\documentclass[aps,prl,twocolumn,groupedaddress]{revtex4-2}
\usepackage{graphicx} 
\usepackage[usenames,dvipsnames]{xcolor} 
\usepackage{tikz}	
\usepackage{tikz-3dplot}
\usepackage{amsthm}
\usepackage{float}
\usepackage{bbm, dsfont}
\usepackage{mathtools}
\usepackage{hyperref}
\usepackage{cleveref}
\usepackage{orcidlink}
\usepackage{color}
\usepackage{colortbl}
\usepackage{tcolorbox}
\usepackage{enumerate}
\usepackage{fontawesome}
\usepackage{adjustbox}
\usepackage{tcolorbox}
\usepackage[T1]{fontenc}
\usepackage{newtxtext,newtxmath}
\usepackage{cancel}
\usepackage{enumitem}
\usepackage{nicematrix}
\usepackage{bm}
\usepackage{mleftright}
\usepackage{array}
\usepackage{physics}
\usepackage{cleveref}
\usepackage{booktabs}

\hypersetup{
    colorlinks=true,%
    allcolors={blue}
}

\usepackage{titlesec}

\titlespacing*{\section}{0.4pt}{0.2\baselineskip}{0.3\baselineskip}

\titleformat{\section}
{\centering\bfseries} 
{\thesection} 
{1em} 
{}

\renewcommand{\thesection}{\arabic{section}}

\newcommand{\centergraphicx}[2]{%
  $\vcenter{\hbox{\includegraphics[width=#1]{#2}}}$%
}

\newcommand{\rmd}{\mathrm{d}}
\newcommand{\rmi}{\mathrm{i}}
\newcommand{\rme}{\mathrm{e}}
\newcommand{\dlog}{\rmd \! \log}
\newcommand{\de}{\rmd \mathcal{E}}

\begin{document}


\title{From $\dlog$ to $\de$: Canonical Elliptic Integrands and Modular Symbol Letters with Pure eMPLs}


\author{Li Lin Yang$\orcidlink{0000-0001-7707-8138}$}
\email{yanglilin@zju.edu.cn}
\affiliation{
    Zhejiang Institute of Modern Physics, 
    School of Physics, Zhejiang University,
    Hangzhou 310058, China
}

\author{Yiyang Zhang$\orcidlink{0009-0005-3972-2611}$}
\email{yiyangzhang@zju.edu.cn}
\affiliation{
    Zhejiang Institute of Modern Physics, 
    School of Physics, Zhejiang University, 
    Hangzhou 310058, China
}


\date{\today}

\begin{abstract}
We propose `$\de$-forms’ as fundamental building blocks of canonical integrands for elliptic Feynman integrals, which lead to Kronecker-Eisenstein $\omega$-form symbol letters. Built upon pure elliptic multiple polylogarithms, they provide a natural extension of the `$\dlog$-form’ integrands and $\dlog$ letters for polylogarithmic cases. By introducing an extended basis treating all marked points equally, we manifest a hidden symmetry structure in the canonical connection matrix, and demonstrate its covariance under modular transformations. Our result provides a novel perspective on describing canonical bases and symbol letters in a unified language of pure functions.
\end{abstract}


\maketitle

\allowdisplaybreaks

\section{Introduction and summary}

\label{sec:intro}

Driven by increasingly precise experimental data from colliders, we have entered the era of precision measurements. To match the experimental accuracy, theoretical predictions at high perturbative orders in quantum field theories are required, where Feynman integrals play a ubiquitous role. Any integrals in an integral family can be expressed as linear combinations of a finite number of \emph{master integrals} (MIs), often found with \emph{integration-by-parts} (IBP) relations~\cite{Tkachov:1981wb,Chetyrkin:1981qh,Laporta:2000dsw}. In order to evaluate masters integrals analytically, the \emph{differential equations} method~\cite{Kotikov:1990kg,Kotikov:1991pm,Gehrmann:1999as}, especially the \emph{canonical differential equations} method~\cite{Henn:2013pwa}, has been widely applied with great success. If a canonical basis is found, the master integrals can be expressed in terms of \emph{iterated integrals}~\cite{Chen:1977oja} straightforwardly. The simplest kind of iterated integrals are \emph{multiple polylogarithms} (MPLs)~\cite{Goncharov:1998kja,Goncharov:2001iea}. For Feynman integral families associated with MPLs, canonical bases can be derived using algorithms based on differential equations~\cite{Lee:2014ioa,Lee:2017oca,Gituliar:2017vzm,Prausa:2017ltv,Meyer:2017joq,Lee:2020zfb,Dlapa:2020cwj} or directly constructed with \emph{$\dlog$-forms} integrands~\cite{Henn:2020lye,Chen:2020uyk,Chen:2022lzr}. The construction is often carried out through cut-based analysis~\cite{Cachazo:2008vp,Arkani-Hamed:2010pyv,Primo:2016ebd,Frellesvig:2017aai,Bosma:2017ens,Primo:2017ipr,Harley:2017qut} in \emph{(loop-by-loop) Baikov representations}~\cite{Baikov:1996rk,Baikov:1996iu,Frellesvig:2017aai,Harley:2017qut}, which can be described in the language of \emph{twisted cohomology}~\cite{Mastrolia:2018uzb,Frellesvig:2019uqt}. However, starting from two loops, many Feynman integrals go beyond MPLs~\cite{Bourjaily:2022bwx}. Thus, in order to improve theoretical precision, it is of great importance to extend the understanding of canonical bases to \emph{elliptic Feynman integrals} and beyond.

In the last decade, there are huge advancements in understanding the underlying geometries of Feynman integrals, and various algorithms have been proposed to find canonical bases for elliptic families and beyond~\cite{Adams:2018yfj,Adams:2018bsn,Giroux:2022wav,Dlapa:2022wdu,Pogel:2022vat,Gorges:2023zgv,Duhr:2025lbz,Maggio:2025jel,e-collaboration:2025frv,Duhr:2025xyy,Bree:2025tug}. These advancements enable a large number of phenomenological calculations~\cite{Bogner:2019lfa,Muller:2022gec,Pogel:2022yat,Pogel:2022ken,Jiang:2023jmk,Delto:2023kqv,Giroux:2024yxu,Duhr:2024bzt,Wang:2024ilc,Forner:2024ojj,Schwanemann:2024kbg,Marzucca:2025eak,Becchetti:2025rrz,Becchetti:2025oyb,Pogel:2025bca,Duhr:2025kkq}. Besides these differential equation based methods, there are also integrand-level approaches in Refs.~\cite{Frellesvig:2021hkr,Chaubey:2025adn}, which however do not directly lead to \emph{uniform transcendentality} (UT) or \emph{$\varepsilon$-factorization}. In Ref.~\cite{Chen:2025hzq}, a generic form of canonical univariate elliptic integrands has been derived, which can be applied to a wide range of elliptic Feynman integrals. However, the integrands presented in that work are rather complicated, and it is difficult to intuitively understand why they correspond to a canonical basis. Moreover, even with canonical integrals at hand, it is still important to study their \emph{symbol letters}. These are much less investigated~\cite{Broedel:2018iwv,Broedel:2018qkq,Broedel:2018rwm,Broedel:2019hyg,Duhr:2019rrs,Weinzierl:2020fyx,Wilhelm:2022wow}, and most of them are focused on banana families~\cite{Adams:2017ejb,Broedel:2019kmn,Broedel:2021zij,Pogel:2022yat,Duhr:2025tdf,Duhr:2025ppd,Duhr:2025ouy}. Therefore, the structure of canonical integrands and the associated symbol letters still calls for better understanding.

Canonical bases are intimately related to the concept of \emph{pure functions}~\cite{Cachazo:2008vp,Arkani-Hamed:2010pyv,Henn:2013pwa,Broedel:2018qkq,Frellesvig:2023iwr}. A definition of pure functions is given in Ref.~\cite{Broedel:2018qkq}: A pure function is \emph{unipotent} and its total differential involves only pure functions and one-forms with at most \emph{logarithmic singularities}. In polylogarithmic cases, MPLs are pure functions themselves, whose integrands and total differentials both consist of $\dlog$-forms. There are attempts to extend $\dlog$-forms construction with pure \emph{elliptic multiple polylogarithms} (eMPLs)~\cite{Gorges:2023zgv,Duhr:2025lbz,Chaubey:2025adn}, while they rely on information of differential equations or give up $\varepsilon$-factorization in the end. A common obstacle lies on the fact that a complete basis of Feynman integrals requires an integrand in the form of the \emph{Abelian differential} of the second kind, which has a double pole. On the other hand, the integrands of pure eMPLs have at most simple poles. The contradiction prevents the construction of canonical integrands with pure eMPLs straightforwardly.

In this Letter, we propose to use so-called \emph{$\de$-forms} to construct canonical elliptic integrands. They are (linear combinations of) the integrands of pure eMPLs. We derive the corresponding symbol letters using a novel method, and find that they can be naturally expressed in terms of \emph{Kronecker-Eisenstein $\omega$-forms} which appear in the total differentials of pure eMPLs. An important ingredient in our derivation of the canonical integrands is that we successfully find linear combinations of $\de$-forms that can be converted to Feynman integrals via IBP. The resulting basis is canonical as expected, satisfying \emph{$\varepsilon$-factorized} differential equations with at most simple poles, \emph{IBP-equivalent} to $\dlog$-forms integrands in degenerated limits, and has a constant \emph{intersection matrix}, as advocated in the literature~\cite{Dlapa:2022wdu,Frellesvig:2023iwr,Duhr:2024uid}. By considering an extended basis to treat all marked points equally, we reveal a symmetry structure hidden in the canonical connection matrix. 
We also derive the additional fiber transformation for the canonical basis under \emph{modular transformations}, and check the covariance of the corresponding connection matrix and modular symbol letters. These nice properties allow us to apply our results to different kinematic regions in a unified way, and also enable efficient numeric evaluations of the relevant functions~\cite{Weinzierl:2020fyx}. Our approach provides a new perspective on the canonical bases and the symbol letters of polylogarithmic and elliptic integral families, which can be described with a unified language of pure functions.

\section{Setup}

\label{sec:basics}

Without loss of generality, we consider a univariate elliptic integral family on the \emph{maximal cut}~\cite{Primo:2016ebd,Frellesvig:2017aai,Bosma:2017ens,Primo:2017ipr,Harley:2017qut}, where the elliptic curve is in the \emph{Legendre normal form}. The results can be applied to generic families via suitable \emph{M\"obius transformations}~\cite{Chen:2025hzq}. An integral in this family can be expressed as 
\begin{equation}
    \label{eq:int}
    I 
    = \int \limits_\mathcal{C} \, u_L (x) \, \varphi
    = \int \limits_\mathcal{C} \, \bar{u}_L (x) \frac{\varphi}{\sqrt{P_L (x)}}
    = \int \limits_\mathcal{C} \, \bar{u}_L (x) \, \phi
    \, ,
\end{equation}
where the \emph{twist} is given by
\begin{equation}
    \label{eq:twist}
    u_L (x)
    = {[P_L (x)]}^{- 1 / 2} \bar{u}_L (x) 
    = {[P_L (x)]}^{- 1 / 2} \prod_{i = 2}^{n + 1} {(x - e_i)}^{- \beta_i \varepsilon}
    \, .
\end{equation}
The Legendre elliptic curve is defined by a cubic polynomial  
\begin{equation}
    \label{eq:ellcurve}
    {y}^2
    = P_L (x) 
    = (x - e_2) (x - e_3) (x - e_4)
    = x (x - \lambda) (x - 1)
    \, ,
\end{equation}
where we adopt the choice of branch cuts in Ref.~\cite{Bogner:2017vim}
\begin{equation}
    \label{eq:branch}
    y 
    = \sqrt{P_L (x)}
    = - \sqrt{x} \sqrt{x - \lambda} \sqrt{x - 1}
    \, .
\end{equation}

It is well-known that there is an \emph{isomorphism}, the \emph{Abel's map}, between an elliptic curve and a torus. It is defined in terms of the Abelian differential of the first kind:
\begin{equation}
    \label{eq:abelmap}
    \mleft(x, \pm \sqrt{P_L (x)}\mright) 
    \mapsto z 
    = \pm \int \limits_{e_1 = \infty}^{x} \, \frac{\rmd x^\prime}{\sqrt{P_L (x^\prime)}} \quad {\mathrm{mod}} \; \Lambda 
    \, ,
\end{equation}
where $\Lambda = \mathbb{Z} \omega_1 + \mathbb{Z} \omega_2$ is the lattice generated by two periods 
\begin{equation}
    \label{eq:periods}
    \omega_1
    = 2 \int \limits_{e_1 = \infty}^{e_4 = 1} \, \frac{\rmd x}{\sqrt{P_L (x)}}
    \, ,
    \quad
    \omega_2 
    = 2 \int \limits_{e_1 = \infty}^{e_2 = 0} \, \frac{\rmd x}{\sqrt{P_L (x)}}
    \, .
\end{equation}
We mod out $\Lambda$ in Eq.~\eqref{eq:abelmap} to get rid of the multi-valuedness and ensure an isomorphism. There is freedom in choosing the generators of the lattice, which are related by ${\rm SL} (2, \mathbb{Z})$ modular transformations
\begin{equation}
    \label{eq:modtrans}
    \begin{pmatrix}
        \omega_2
        \\
        \omega_1
    \end{pmatrix}
    \mapsto 
    \begin{pmatrix}
        \omega_2^\prime
        \\
        \omega_1^\prime
    \end{pmatrix}
    = 
    \begin{pmatrix}
        a & b 
        \\
        c & d
    \end{pmatrix}
    \,
    \begin{pmatrix}
        \omega_2
        \\
        \omega_1
    \end{pmatrix}
    \, ,
    \quad 
    \begin{pmatrix}
        a & b 
        \\
        c & d
    \end{pmatrix} 
    \in {\rm SL} (2,\mathbb{Z})
    \, .
\end{equation}

We are also free to rescale the holomorphic differential by a non-zero factor without breaking the isomorphism. With a factor $1 / \omega_1$, the lattice turns into $\Lambda_\tau = \mathbb{Z} + \mathbb{Z} \tau$, and marked points $\mleft(e_i, \pm \sqrt{P_L (e_i)}\mright)$ on the curve are mapped to $z_i^\pm$:
\begin{equation}
    \label{eq:modvar}
    \tau 
    = \frac{\omega_2}{\omega_1}
    \in \mathbb{H}
    \, ,
    \;
    z_i^\pm
    = \pm \frac{1}{\omega_1} \int \limits_{e_1 = \infty}^{e_i} \, \frac{\rmd x^\prime}{\sqrt{P_L (x^\prime)}} 
    \, ,
    \;
    (i = 1, \cdots, n + 1)
    \, .
\end{equation}
Specifically, the branch points of the curve are mapped to rational points on the torus 
\begin{equation}
    \label{eq:branchpointzi} 
    z_1 
    = 0
    \, ,
    \quad 
    z_2
    = \frac{1}{2} \tau
    \, ,
    \quad 
    z_3 
    = \frac{1}{2} + \frac{1}{2} \tau
    \, ,
    \quad 
    z_4
    = \frac{1}{2}
    \, .
\end{equation} 
$(\tau, z_i)$ with $i = 5, \cdots, n + 1$ are \emph{modular variables} on the torus. They are functions of the variables $(\lambda, e_i)$ on the curve, and the underlying isomorphism allows us to invert the relations. Under modular transformations, they transform as 
\begin{equation}
    \label{eq:modtransmod}
    \tau 
    \mapsto \tau^\prime
    = \frac{a \tau + b}{c \tau + d}
    \, ,
    \quad 
    z_i 
    \mapsto z_i^\prime
    = \frac{z_i}{c \tau + d}
    \, .
\end{equation}

There are different definitions of pure eMPLs in the literature. One of them defines the pure eMPLs $\widetilde{\Gamma}$ as iterated integrals on the torus \cite{Broedel:2017kkb}:
\begin{equation}
    \label{eq:emplgamma}
    \widetilde{\Gamma} \mleft(
        \begin{smallmatrix}
            n_1 & \cdots & n_k 
            \\
            z_1 & \cdots & z_k
        \end{smallmatrix}; 
        z, \tau
    \mright)
    = \int \limits_{0}^{z} \, \rmd z^\prime \, g^{(n_1)} (z^\prime - z_1, \tau) \, \widetilde{\Gamma} \mleft(
        \begin{smallmatrix}
            n_2 & \cdots & n_k
            \\
            z_2 & \cdots & z_k
        \end{smallmatrix};
        z^\prime, \tau
    \mright)
    \, ,
\end{equation}
which is of \emph{length}-$k$ and \emph{transcendental weight}-$\sum_{i = 1}^{k} n_k$. Functions $g^{(k)}$ are generating series of \emph{Kronecker-Eisenstein series}~\cite{Broedel:2018iwv}, and $g^{(k)} (z, \tau)$ has at most simple poles in $z$.
The total differential of a length-$k$ $\widetilde{\Gamma}$ is a linear combination of length-$(k - 1)$ $\widetilde{\Gamma}$ multiplied by pure numbers and Kronecker-Eisenstein $\omega$-forms~\cite{Broedel:2018iwv,Broedel:2018qkq}. An $\omega$-form of \emph{(quasi-)modular weight}-$(k - 2)$ is defined as
\begin{equation}
    \label{eq:KEforms}
    \omega^{(k)} (z, \tau)
    = \frac{1}{{\pi}^{k - 2}} \mleft[(k - 1) g^{(k)} (z, \tau) \frac{\rmd \tau}{2 \pi \rmi} + g^{(k - 1)} (z, \tau) \rmd z \mright]
    \, .
\end{equation}
Thus, the $\widetilde{\Gamma}$ functions satisfy the definition of pure functions, at least for $\tau$ on the \emph{complex upper half-plane} $\mathbb{H}$ and at the \emph{cusp} $\rmi \infty$~\footnote{It was suggested in Ref.~\cite{Frellesvig:2023iwr} that they only locally (for $\tau$ in a patch containing the cusp $\rmi \infty$) satisfy the definition of pure functions in Ref.~\cite{Broedel:2018qkq}, since an $\omega$-form may have an order-$k$ pole in $\tau$ at a finite cusp. This is not a problem in practice, since we can cover the full kinematic space with such local patches with the help of modular transformations.}.

Another convenient class of pure eMPLs, denoted by $\mathcal{E}$, can be expressed as constant linear combinations of $\widetilde{\Gamma}$. These functions possess \emph{definite parity} and manifestly contain MPLs~\cite{Broedel:2018qkq} (note that the definition in Ref.~\cite{Broedel:2018qkq} employs a quartic curve). The definition is 
\begin{equation}
    \label{eq:pureE}
    \mathcal{E} \mleft(
        \begin{smallmatrix}
            n_1 & \cdots & n_k 
            \\
            e_1 & \cdots & e_k
        \end{smallmatrix}; 
        x, \lambda
    \mright)
    = \int \limits_{0}^{x} \, \rmd x^\prime \, \Psi_{n_1} (e_1, x^\prime, \lambda) \, \mathcal{E} \mleft(
        \begin{smallmatrix}
            n_2 & \cdots & n_k
            \\
            e_2 & \cdots & e_k
        \end{smallmatrix};
        x^\prime, \lambda
    \mright)
    \, ,
\end{equation}
where the integration kernels are 
\begin{subequations}
    \label{eq:purekernel}
    \begin{align}
        & \de \mleft(
            \begin{smallmatrix}
                0 
                \\
                0
            \end{smallmatrix}; 
            x, \lambda
            \mright)
        = \Psi_{0} (0, x) \, \rmd x
        = \rmd z 
        = \rmd \widetilde{\Gamma} \mleft(
            \begin{smallmatrix}
                0 
                \\
                0
            \end{smallmatrix}; 
            z, \tau
        \mright)
        \, ,
        \\
        & \de \mleft(
            \begin{smallmatrix}
                \pm n 
                \\
                e_i
            \end{smallmatrix}; 
            x, \lambda
            \mright)
        = \Psi_{\pm n} (e_i, x) \, \rmd x
        \nonumber
        \\
        & = \mleft[g^{(n)} (z - z_i, \tau) \pm g^{(n)} (z + z_i, \tau) - 2 \delta_{\pm n, 1} g^{(1)} (z, \tau)\mright] \, \rmd z
        \nonumber
        \\
        & = \rmd \mleft[\widetilde{\Gamma} \mleft(
            \begin{smallmatrix}
                n 
                \\
                z_i^+
            \end{smallmatrix}; 
            z, \tau
        \mright) \pm \widetilde{\Gamma} \mleft(
            \begin{smallmatrix}
                n 
                \\
                z_i^-
            \end{smallmatrix}; 
            z, \tau
        \mright) - 2 \delta_{\pm n, 1} \widetilde{\Gamma} \mleft(
            \begin{smallmatrix}
                1 
                \\
                0
            \end{smallmatrix}; 
            z, \tau
        \mright)\mright]
        \, ,
        \; 
        (n > 0)
        \, .
    \end{align}
\end{subequations}

\section{Canonical elliptic integrands and symbol letters}

\label{sec:can}

We now introduce $\de$-forms relevant for canonical Feynman integrals. For that purpose, we need to find linear combinations that can be related to the three kinds of Abelian differentials. The ones associated with Abelian differentials of the first and third kinds can be easily found. However, the second kind of Abelian differential has a double pole, while $\de$-forms have at most simple poles. In addition, the double pole cannot be removed with IBP relations induced by rational functions due to the irreducibility of Abelian differentials. This obvious contradiction prevents us from constructing an integrand of the second kind exactly in terms of $\de$-forms. Consequently, we must take more generalized IBP relations induced by transcendental functions into account. This is precisely how integration kernels of eMPLs were constructed in Ref.~\cite{Broedel:2017kkb}. With the help of the generalized IBP relations (extended to the twisted case), we find a specific constant linear combination of $\de$-forms with a uniform modular weight-$2$, which is IBP-equivalent to an integrand expressed with the three kinds of Abelian differentials. Our findings can be summarized by the following expressions of the integrands in terms of $\de$-forms:
\begin{subequations}
    \label{eq:canint}
    \begin{align}
        \phi_1 
        & = \pi \, \de \mleft(
            \begin{smallmatrix}
                0 
                \\
                0
            \end{smallmatrix}; 
            x, \lambda
            \mright)
        \, ,
        \\
        \phi_{i - 3}
        & = \beta_i \, \de \mleft(
            \begin{smallmatrix}
                - 1 
                \\
                e_i
            \end{smallmatrix}; 
            x, \lambda
            \mright)
        \, ,
        \quad 
        (i = 5, \cdots, n + 1)
        \, ,
        \\
        \phi_{n - 1}
        & = \frac{1}{\pi} \sum_{i = 1}^{n + 1} \beta_i \, \de \mleft(
            \begin{smallmatrix}
                2 
                \\
                e_i
            \end{smallmatrix}; 
            x, \lambda
            \mright)
        \, ,
    \end{align}
\end{subequations}
where the exponents $\beta_i$ are defined in Eq.~\eqref{eq:twist}, and we define $\beta_1 = - \sum_{i = 2}^{n + 1} \beta_i$. The factors $\pi$ and $1 / \pi$ are introduced to adjust the transcendental weights of the differentials to be $1$. The above expressions intrinsically encode rich information about the integrals without reference to differential equations. They provide us with integrand-level understanding of the basis and facilitate systematic analysis of the associated symbol letters, as well as their behaviors under modular transformations, as we will see soon.

Starting from the integrands, we can derive the differential equations for the corresponding integrals $\vec{M}$. The connection matrix is $\varepsilon$-factorized with at most simple poles as expected:
\begin{equation}
    \label{eq:cande}
    \rmd \vec{M}
    = \varepsilon \bm{A} \, \vec{M}
    \, .
\end{equation}
In addition, the integrands are IBP-equivalent to $\dlog$-forms in the degenerated limit $\lambda \to 0$, and the basis has a constant intersection matrix. All the above are properties of a canonical basis as advocated in the literature~\cite{Dlapa:2022wdu,Frellesvig:2023iwr,Duhr:2024uid}.

The structure of the canonical integrands in Eq.~\eqref{eq:canint} seems to suggest that the entries of the connection matrix should be expressed with $\omega$-forms of uniform modular weights. However, it turns out that the entries in the first columns are of mixed modular weights. Most of the mixing behaviors can be made manifest by taking \emph{self-duality} relations~\cite{Pogel:2024sdi,Duhr:2024xsy} into account, with which the number of independent entries reduces to $[(n - 1) n] / 2$. For example, we have self-duality relations 
\begin{equation}
    \label{eq:selfdualityeg}
    {\bm A}_{i - 3, 1}
    = - 2 \beta_i [2 (\beta_2 + \beta_3) {\color{orange} {\bm A}_{1, i - 3}} + {\color{blue} {\bm A}_{n - 1, i - 3}}]
    \, ,
\end{equation}
where $i = 5, \cdots, n + 1$. Based on the modular weights of canonical integrands in Eq.~\eqref{eq:canint}, we expect ${\color{orange} {\bm A}_{1, i - 3}}$ and ${\color{blue} {\bm A}_{n - 1, i - 3}}$ to be of modular weight-$(- 1)$ and $1$ respectively. Note that here and in the following, we show functions of different modular weights in distinct colors. We can express ${\color{orange} {\bm A}_{1, i - 3}}$ and ${\color{blue} {\bm A}_{n - 1, i - 3}}$ in $\omega$-forms with \emph{numeric bootstrap}
\begin{subequations}
    \label{eq:letterinomegaeg}
    \begin{align}
        {\color{orange} {\bm A}_{1, i - 3}}
        & = {\color{orange} \omega^{(1)} (z_i, \tau)}
        \, ,
        \\ 
        {\color{blue} {\bm A}_{n - 1, i - 3}}
        & = \sum_{j = 1}^{n + 1} \beta_j \mleft[{\color{blue} \omega^{(3)} (z_i + z_j, \tau)} + {\color{blue} \omega^{(3)} (z_i - z_j, \tau)}\mright]
        \, .
    \end{align}
\end{subequations}
Thus, from Eq.~\eqref{eq:selfdualityeg} and Eq.~\eqref{eq:letterinomegaeg}, we can see that ${\bm A}_{i - 3, 1}$ is of mixed modular weights. Moreover, the appearance of $\beta_2$ and $\beta_3$ in the mixing coefficients seems puzzling.

The phenomenon demonstrated above can be traced back to the integrands in Eq.~\eqref{eq:canint}. In $\phi_{n - 1}$, all the marked points $e_j$'s with $j = 1, \cdots, n + 1$ have equal contribution. However, for $\phi_{i - 3}$, we do not have $\de$-forms involving $e_1, e_2, e_3, e_4$. In fact, these $\de$-forms are either vanishing or equivalent to $\phi_1$ up to a constant factor. In other words, the exchange symmetry of marked points is broken by the requirement of linear independence, which leads to the puzzling behavior of the connection matrix.

In order to reveal the hidden symmetry, we would like to derive a connection matrix with uniform modular weights. For that purpose, it is helpful to consider an extended `basis' which is not linear independent:
\begin{subequations}
    \label{eq:extcanint}
    \begin{align}
        \widetilde{\phi}_0
        & = \pi \, \de \mleft(
            \begin{smallmatrix}
                0 
                \\
                0
            \end{smallmatrix}; 
            x, \lambda
            \mright)
        \, ,
        \\
        \widetilde{\phi}_i
        & = \beta_i \, \de \mleft(
            \begin{smallmatrix}
                - 1 
                \\
                e_i
            \end{smallmatrix}; 
            x, \lambda
            \mright)
        \, ,
        \quad 
        (i = 1, \cdots, n + 1)
        \, ,
        \\
        \widetilde{\phi}_{n + 2}
        & = \frac{1}{\pi} \sum_{i = 1}^{n + 1} \beta_i \, \de \mleft(
            \begin{smallmatrix}
                2 
                \\
                e_i
            \end{smallmatrix}; 
            x, \lambda
            \mright)
        \, .
    \end{align}
\end{subequations}
They are related to the original basis as:
\begin{equation}
    \label{eq:linrels}
    \widetilde{\phi}_0 
    = \phi_1 
    \, ,
    \quad 
    \widetilde{\phi}_{1, 4}
    = 0
    \, ,
    \quad 
    \widetilde{\phi}_{2, 3}
    = 2 \rmi \beta_{2, 3} \, \phi_1 
    \, ,
    \quad 
    \widetilde{\phi}_i 
    = \phi_{i - 3}
    \, ,
\end{equation}
with $i = 5, \cdots, n + 2$. From the above relations, we can get a glimpse of the reason why we have $\beta_2$ and $\beta_3$ in Eq.~\eqref{eq:selfdualityeg}.

However, if we decide to work with the extended basis, it is no longer possible to obtain the entries through numeric bootstrap. First of all, the numeric differential equations are not uniquely determined due to the linear dependence, and there are ambiguities in writing the numeric connection matrix. It is not \emph{a priori} known how to redistribute the numbers to achieve uniform-weight entries. There are also ambiguities associated with the linear relations among modular symbol letters, \emph{e.g.}, $\omega^{(k)} (z_i + z_4) = \omega^{(k)} (z_i - z_4)$. These relations make it more difficult to write the entries in a symmetric way.

We resolve these obstacles by developing a novel and elegant method to derive the connection matrix in terms of modular symbol letters. This method does not rely on the differential equations of the integrals $\vec{M}$, and only makes use of our $\de$-forms integrands in Eq.~\eqref{eq:extcanint}. Since we are dealing with a canonical basis, the corresponding connection matrix is $\varepsilon$-factorized. Thus, the coefficients in the $\varepsilon$-expansion of the canonical integrals encode the information of the connection matrix in a simple way. To be more explicit, we consider the $\varepsilon$-expansion of the extended basis up to order $\varepsilon^1$:
\begin{equation}
    \label{eq:epsexp}
    \int \limits_\mathcal{C} \, \bar{u}_L (x) \, \widetilde{\phi}
    = \int \limits_\mathcal{C} \, \widetilde{\phi} - \sum_{i = 2}^{n + 1} \beta_i \varepsilon \int \limits_\mathcal{C} \, \widetilde{\phi} \log (x - e_i) + {\cal O} (\varepsilon^2)
    \, .
\end{equation}
We exchange the order of $\varepsilon$-expansion and integration above, which is legal only if the boundary points of the integration contour $\mathcal{C}$ are integrable. Fortunately, we can find such a contour, which is $[e_1, e_4]$. We can express the $\varepsilon^0$ and $\varepsilon^1$ terms in terms of pure eMPLs. We take the total differential of the $\varepsilon^1$ terms using the properties of pure eMPLs, which can be written as linear combinations of the $\varepsilon^0$ terms. The symbol letters can then be extracted from the combination coefficients. At this stage, we do not impose the linear relations in Eq.~\eqref{eq:linrels}, or in other words, we treat all the marked points $z_i$'s as generic and independent for $i = 1, \cdots, n + 1$. We stress that the method above needs no \emph{a priori} knowledge of the connection matrix except the $\varepsilon$-factorization, and the derivation is straightforward and efficient.

Using the above method, we successfully express the connection matrix of the extended basis in modular symbol letters, where all entries are of uniform modular weight. The explicit form is
\begin{equation}
    \label{eq:Atilde}
    \begin{aligned}
        & \widetilde{\bm A} =
        \\
        & \begin{pmatrix}
            {\color{Green} \widetilde{\bm A}_{0, 0}} & {\color{orange} \widetilde{\bm A}_{0, 1}} & \cdots & {\color{orange} \widetilde{\bm A}_{0, n + 1}} & {\color{red} \widetilde{\bm A}_{0, n + 2}}
            \\
            {\color{blue} - 2 \beta_1 \widetilde{\bm A}_{n + 2, 1}} & {\color{Green} \widetilde{\bm A}_{1, 1}} & \cdots & {\color{Green} \frac{\beta_1}{\beta_{n + 1}} \widetilde{\bm A}_{n + 1, 1}} & {\color{orange} - 2 \beta_1 \widetilde{\bm A}_{0, 1}}
            \\
            \vdots & \vdots & \ddots & \vdots & \vdots 
            \\
            {\color{blue} - 2 \beta_{n + 1} \widetilde{\bm A}_{n + 2, n + 1}} & {\color{Green} \widetilde{\bm A}_{n + 1, 1}} & \cdots & {\color{Green} \widetilde{\bm A}_{n + 1, n + 1}} & {\color{orange} - 2 \beta_{n + 1} \widetilde{\bm A}_{0, n + 1}}
            \\
            {\color{Purple} \widetilde{\bm A}_{n + 2, 0}} & {\color{blue} \widetilde{\bm A}_{n + 2, 1}} & \cdots & {\color{blue} \widetilde{\bm A}_{n + 2, n + 1}} & {\color{Green} \widetilde{\bm A}_{0, 0}}
        \end{pmatrix}
        \\
        & - {\color{Green} 2 \beta_1 \, \dlog \omega_1 \, {\bm 1}_{n + 2, n + 2}}
        \, ,
    \end{aligned}
\end{equation}
where the last $\dlog$-form letter originates from the shuffle regularization of eMPLs due to potential divergence at boundary points~\cite{Broedel:2017kkb,Wilhelm:2022wow}. We will ignore this term since it can be easily absorbed into the twist by an ${\omega_1}^{2 \beta_1 \varepsilon}$ factor. We then list the expressions of independent entries in terms of $\omega$-forms, where we will keep the dependence on $\tau$ implicit.
\\
{\color{red} \textit{\textbf{Modular weight-$(- 2)$:}}}
\begin{equation}
    \label{eq:letterm2}
    {\color{red} \widetilde{\bm A}_{0,n + 2}} 
    = {\color{red} \omega^{(0)} (0)}
    \, .
\end{equation}
{\color{orange} \textit{\textbf{Modular weight-$(- 1)$:}}}
\begin{equation}
    \label{eq:letterm1}
    {\color{orange} \widetilde{\bm A}_{0, i}}
    = {\color{orange} \omega^{(1)} (z_i)}
    \, ,
    \quad 
    (i = 1, \cdots, n + 1)
    \, .
\end{equation}
{\color{Green} \textit{\textbf{Modular weight-$0$:}}}
\begin{subequations}
    \label{eq:letter0}
    \begin{align}
        & {\color{Green} \widetilde{\bm A}_{0, 0}}
        = 2 \sum_{k = 1}^{n + 1} \beta_k {\color{Green} \omega^{(2)} (z_k)}
        \, ,
        \\
        & {\color{Green} \widetilde{\bm A}_{i, j}}
        = \beta_i \mleft[{\color{Green} \omega^{(2)} (z_i - z_j)} - {\color{Green} \omega^{(2)} (z_i + z_j)}\mright]
        \nonumber
        \\ 
        & + \delta_{i j} \sum_{k = 1}^{n + 1} \beta_k \mleft[2 {\color{Green} \omega^{(2)} (z_k)} - {\color{Green} \omega^{(2)} (z_i - z_k)} - {\color{Green} \omega^{(2)} (z_i + z_k)}\mright]
        \, ,
    \end{align}
\end{subequations}
where $1 \leqslant j \leqslant i \leqslant n + 1$.
\\
{\color{blue} \textit{\textbf{Modular weight-$1$:}}}
\begin{equation}
    \label{eq:letter1}
    {\color{blue} \widetilde{\bm A}_{n + 2, i}}
    = \sum_{j = 1}^{n + 1} \beta_j \mleft[{\color{blue} \omega^{(3)} (z_i + z_j)} + {\color{blue} \omega^{(3)} (z_i - z_j)}\mright]
    \, ,
\end{equation}
where $i = 1, \cdots, n + 1$.
\\
{\color{Purple} \textit{\textbf{Modular weight-$2$:}}}
\begin{equation}
    \label{eq:letter2}
    {\color{Purple} \widetilde{\bm A}_{n + 2, 0}}
    = 2 \sum_{i, j = 1}^{n + 1} \beta_i \beta_j \mleft[{\color{Purple} \omega^{(4)} (z_i - z_j)} + {\color{Purple} \omega^{(4)} (z_i + z_j)}\mright]
    \, .
\end{equation}
If we apply the linear relations in Eq.~\eqref{eq:linrels}, the extended connection matrix $\widetilde{\bm A}$ degenerates back to the original one ${\bm A}$, which provides a cross-check for the derivation above.

It is natural to expect that canonical integrands and symbol letters are covariant under modular transformations, since we have the freedom in choosing the modular variable $\tau$. This freedom allows us to choose an appropriate $\tau$ for a particular kinematic region, such that the corresponding \emph{nome square} $\bar{q}$ is constrained in a rather small radius
\begin{equation}
    \label{eq:smallq}
    |\bar{q}|
    = \mleft| {\rme}^{2 \pi \rmi \tau} \mright|
    \leqslant {\rme}^{-\sqrt{3}\pi}
    \approx 0.0043
    \, .
\end{equation}
The covariance of the integrands and symbol letters allows us to write the solutions as iterated integrals of the same form in different kinematic regions. The smallness of $|\bar{q}|$ then makes it possible to numerically evaluate these iterated integrals efficiently through $\bar{q}$-expansion. This will be helpful for phenomenological applications.

For the above purpose, we need to investigate the behaviors of the canonical basis and the symbol letters in the connection matrix under modular transformations. Starting from our extended basis \eqref{eq:extcanint} and the explicit form of the symbol letters \eqref{eq:letterm2}-\eqref{eq:letter2}, this can be done in a simple and systematic way. Under a modular transformation, we require that the symbol letters are expressed as the same $\omega$-forms, with transformed arguments according to Eq.~\eqref{eq:modtransmod}. To achieve that, it is known that we need to apply a fiber transformation on the basis~\cite{Weinzierl:2020fyx}, which is induced by the modular transformation on the variables in the integrands (note that we need to plug Eq.~\eqref{eq:purekernel} into Eq.~\eqref{eq:extcanint} to express the integrands in terms of modular variables, before applying the transformation). The explicit form of the additional fiber transformation can be found with the help of IBP relations for $\widetilde{\phi}_{n - 2}$, which is given by
\begin{equation}
    \label{eq:fibermodtranscanext}
    \widetilde{{\cal T}}
    = 
    \begin{pmatrix}
        \frac{1}{c \tau + d} & 0 & \cdots & 0 & 0
        \\
        - \frac{4 \rmi c \beta_1 z_1}{c \tau + d} & 1 & \cdots & 0 & 0
        \\
        \vdots & \vdots & \ddots & \vdots & \vdots
        \\
        - \frac{4 \rmi c \beta_{n + 1} z_{n + 1}}{c \tau + d} & 0 & \cdots & 1 & 0
        \\
        - \frac{4 {c}^2 \sum_{i = 1}^{n + 1} \beta_i {z_i}^2}{c \tau + d} + \frac{2 \rmi c}{\pi \varepsilon} & - 2 \rmi c z_1 & \cdots & - 2 \rmi c z_{n + 1} & c \tau + d 
    \end{pmatrix}
    \, . 
\end{equation}
The extended basis after the transformation remains canonical, and satisfies similar linear relations as Eq.~\eqref{eq:linrels}. These relations can be used to obtain a linearly independent basis and its associated connection matrix.

The results outlined in this Section is generic enough and can be readily applied (with suitable M\"obius transformations) to any univariate elliptic families with an arbitrary number of scales. We have tested our results in various examples with success. These examples are summarized in Tab.~\ref{tab:eg}. In particular, we have obtained new results for an integral family that were not reported in the literature.

\begin{table}[ht]
    \centering
    \begin{tabular}{*4c}
        \toprule
        \textbf{Diagram} & \textbf{Dimensions} & \textbf{Scales} & \textbf{References} 
        \\
        \midrule
        \centergraphicx{1.3cm}{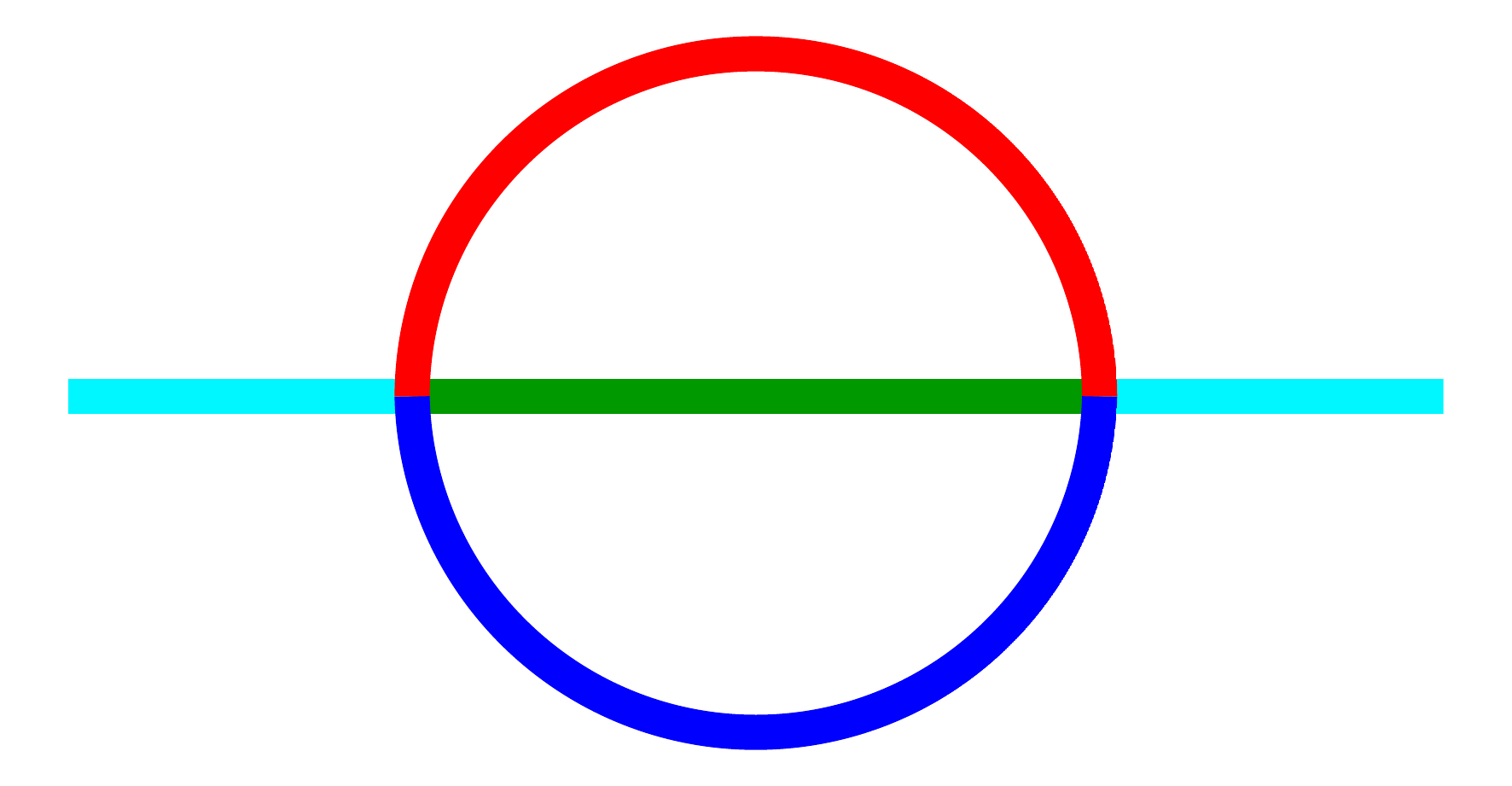} & $4$ & $4$ & \cite{Bogner:2019lfa}
        \\ 
        \centergraphicx{1.3cm}{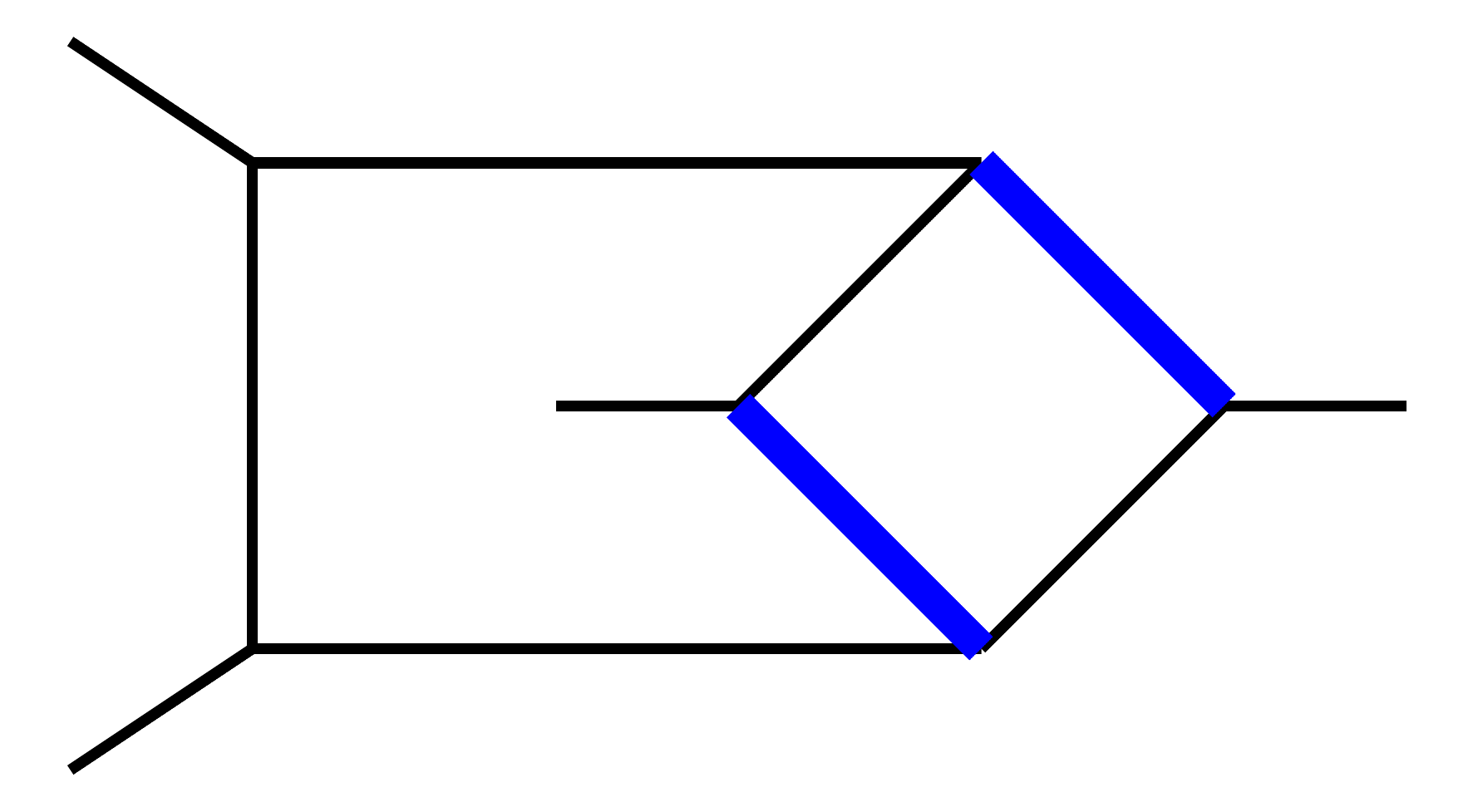} & $4$ & $3$ & \cite{Schwanemann:2024kbg}
        \\
        \centergraphicx{1.3cm}{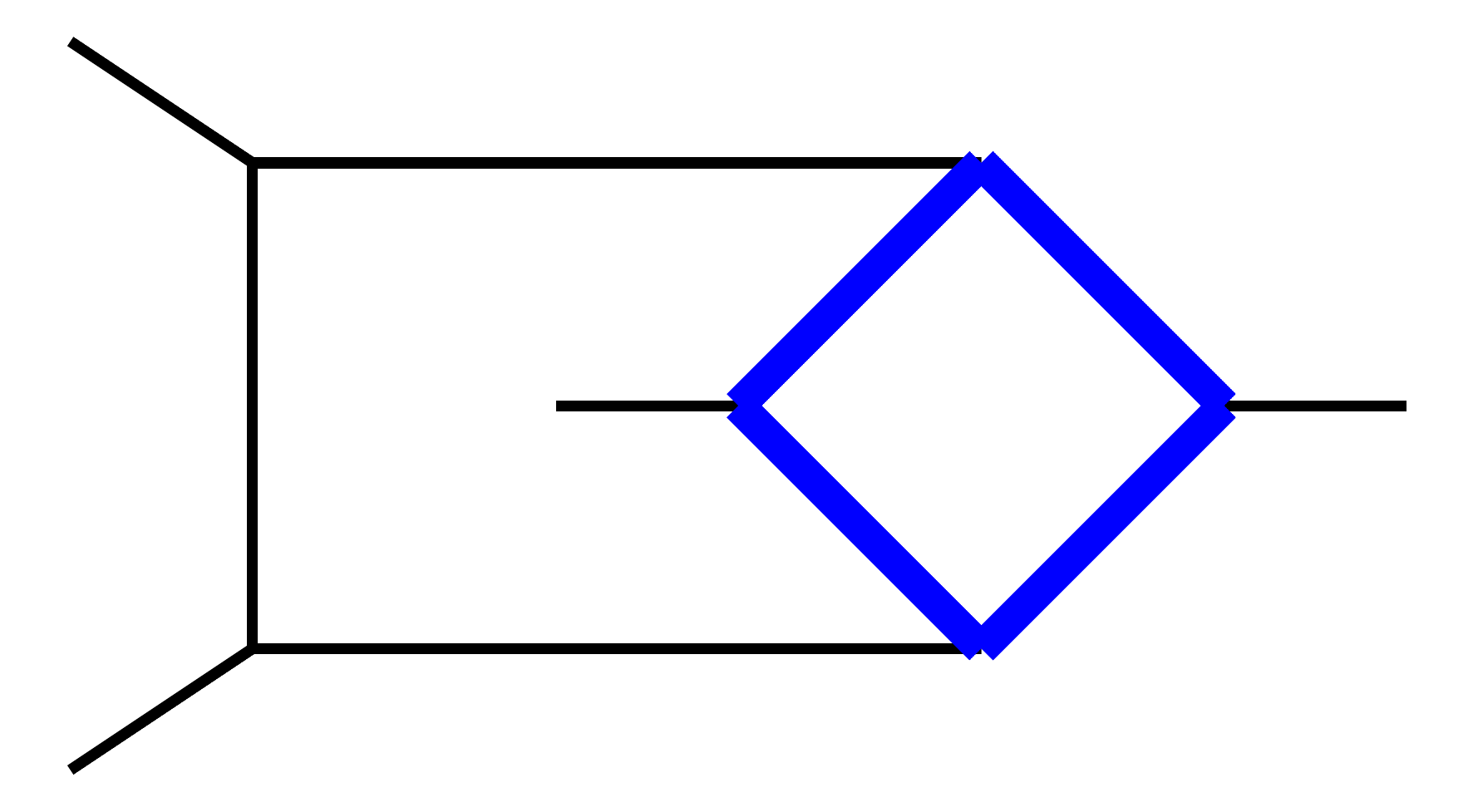} & $4$ & $3$ & \cite{Becchetti:2025rrz}
        \\
        \centergraphicx{1.3cm}{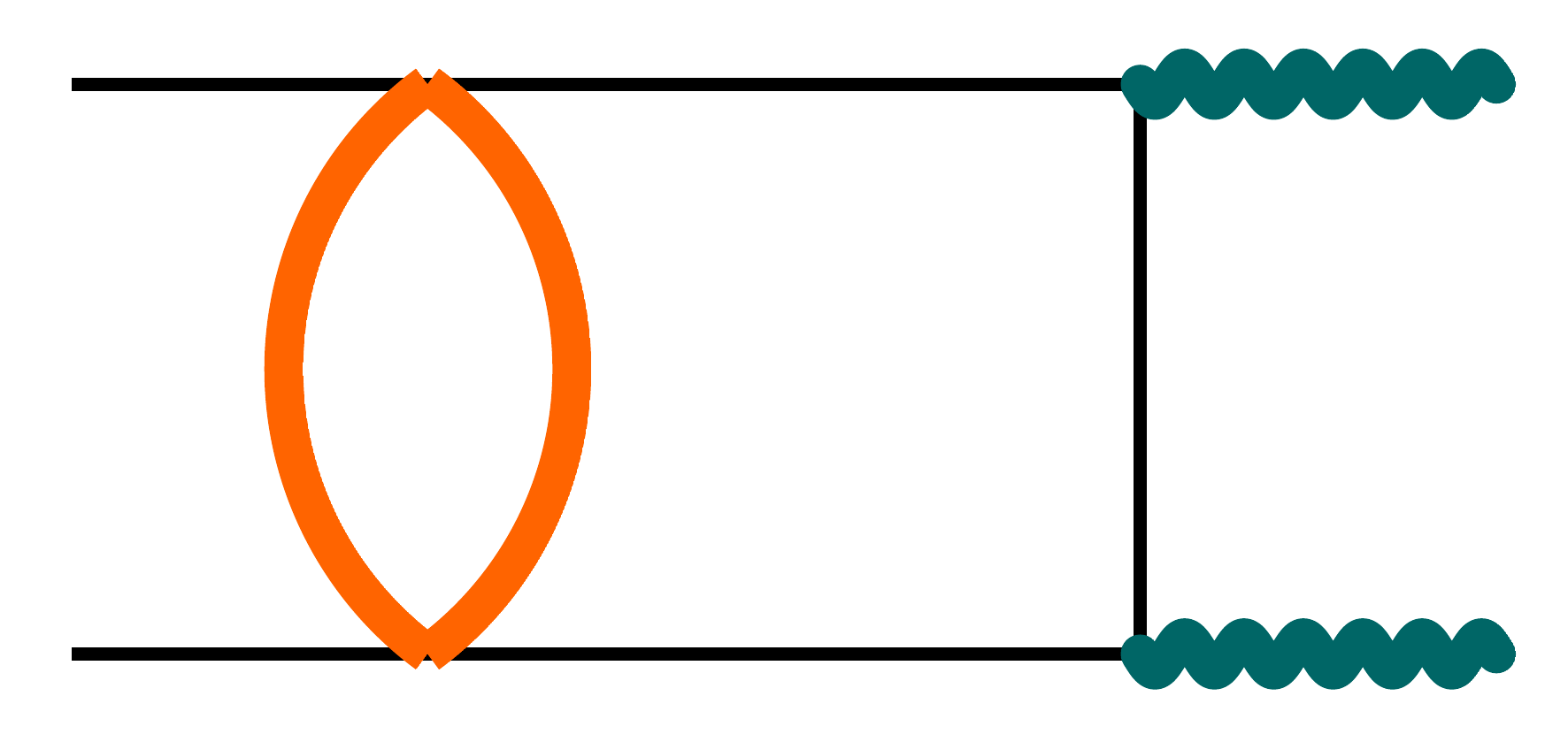} & $3$ & $4$ & \cite{Chen:2025hzq}
        \\
        \centergraphicx{1.3cm}{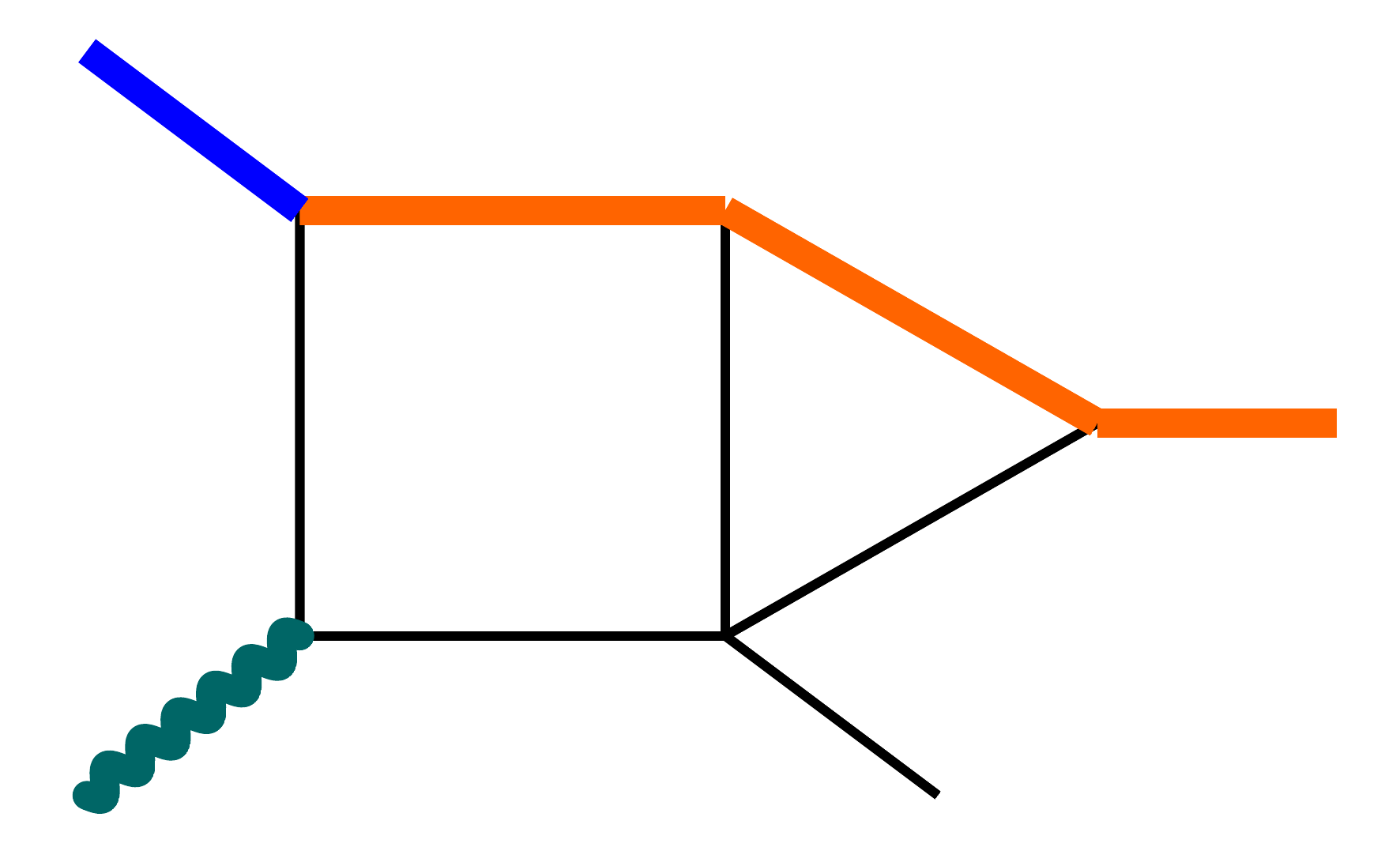} & $3$ & $5$ & \cite{Chen:2025hzq}
        \\
        \centergraphicx{1.3cm}{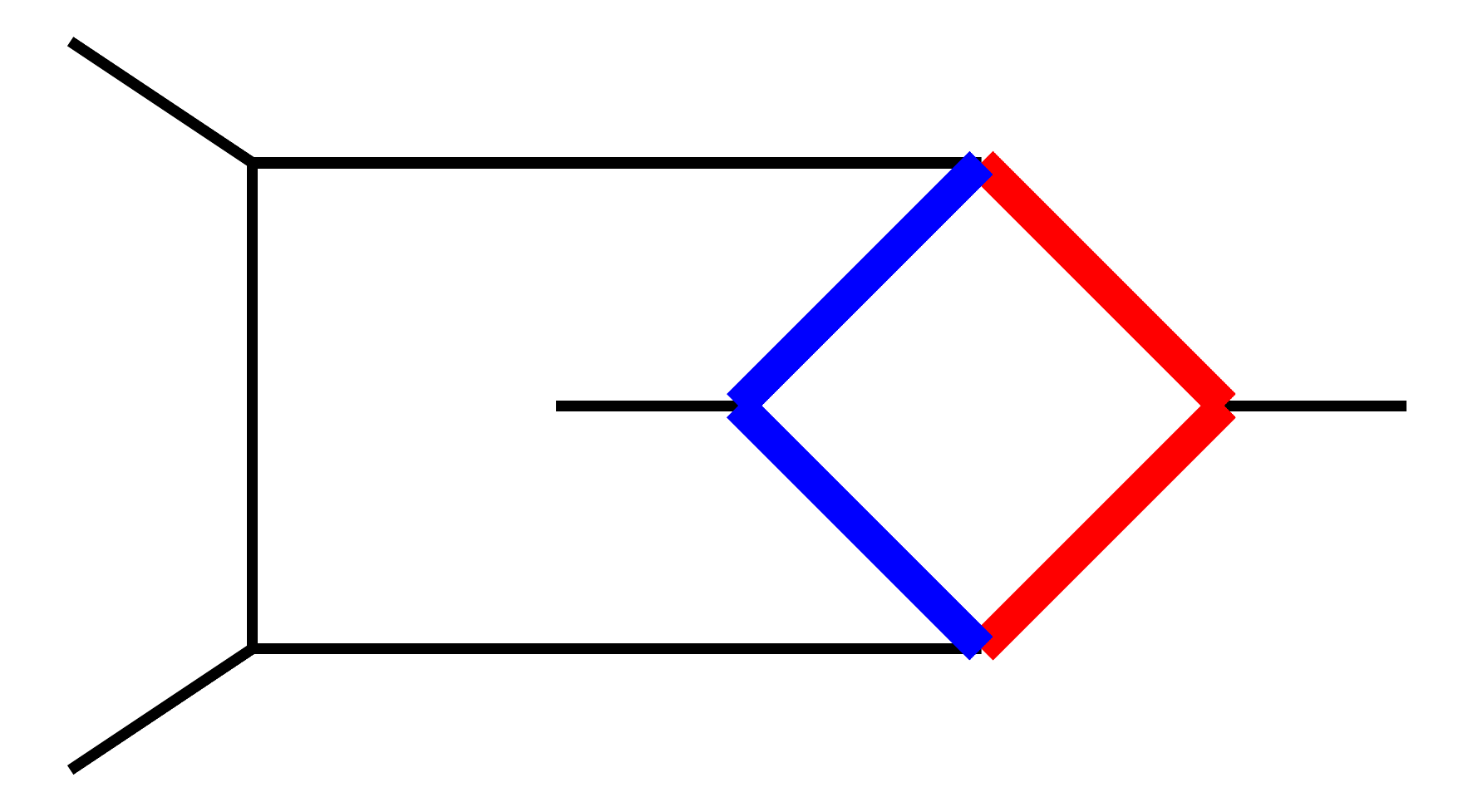} & $6$ & $4$ & New
        \\
        \bottomrule
    \end{tabular}
    \caption{Elliptic integral families to which we have applied our results. Thick lines are massive propagators, where distinct colors represent different masses. For each family, the number of master integrals in the top sector (the dimension), the number of kinematic scales, and relevant references are listed. Notably, the last family with $6$ master integrals represents a new result in this Letter.}
  \label{tab:eg}
\end{table}

For the sunrise family in the first row of Tab.~\ref{tab:eg}, there are also discussions about covariant modular symbol letters in the literature \cite{Weinzierl:2020fyx}. The results were expressed in terms of $\omega$-forms with more complicated arguments and coefficients. Our approach provides much more simplified expressions, thanks to our pure eMPL integrands \eqref{eq:extcanint} and our new method for directly deriving the entries in the connection matrix. Our symbol letters in that family are expressed exclusively in terms of $\omega$-forms with simple arguments $z_i \pm z_j$ for $1 \leqslant j \leqslant i \leqslant n + 1$, and are covariant under modular transformations.

In view of the results presented in this Letter, we now have a unified language of pure functions to describe the canonical integrands and the symbol letters for both polylogarithmic and elliptic Feynman integrals. This is summarized in Tab.~\ref{tab:pure}.
\begin{table}[h]
    \centering
    \begin{tabular}{*4c}
        \toprule
        \textbf{Geometry} & \textbf{Pure functions} & \textbf{Integrands} & \textbf{Letters} 
        \\
        \midrule
        Riemann sphere 
        & $G (e_1, \cdots, e_n; x)$
        & $\dlog (x - e_i)$ 
        & $\dlog (e_i - e_j)$
        \\
        Elliptic curve
        & $\mathcal{E} \mleft(
            \begin{smallmatrix}
                n_1 & \cdots & n_k 
                \\
                e_1 & \cdots & e_k
            \end{smallmatrix}; 
            x, \lambda
        \mright)$
        & $\de \mleft(
            \begin{smallmatrix}
                n_i  
                \\
                e_i
            \end{smallmatrix}; 
            x, \lambda
        \mright)$
        & $\omega^{(m)} (z_i - z_j, \tau)$
        \\
        \bottomrule
    \end{tabular}
    \caption{The building blocks for canonical integrands and symbol letters in terms of pure functions for univariate polylogarithmic and elliptic integral families.}
  \label{tab:pure}
\end{table}

\section{Conclusion and outlook}

\label{sec:conclusion}

In this Letter, we propose $\de$-forms as fundamental building blocks of canonical integrands for elliptic Feynman integrals. They are constructed from pure eMPLs, and can be converted to Feynman integrands through non-trivial IBP relations. The resulting symbol letters can be expressed in terms of modular covariant $\omega$-forms appearing in the total differentials of pure eMPLs. The $\de$-forms and the $\omega$-form symbol letters serve as the natural elliptic counterparts to the $\dlog$-forms and $\dlog$ letters for polylogarithmic integral families. To our knowledge, the $\de$-forms construction of canonical integrands is the first purely integrand-level construction for generic univariate elliptic families, that automatically leads to UT and $\varepsilon$-factorization without manipulating the differential equations. The generic $\omega$-form symbol letters are derived for the first time using a novel method developed in this Letter. We successfully apply our method to various examples including a new result not previously available in the literature.

By employing an extended basis that treats all marked points equally, we reveal a hidden symmetry structure in the canonical connection matrix, which has not been previously reported. The extended connection matrix consists of entries with uniform modular weights, and transforms covariantly under modular transformations with a suitable fiber transformation. These properties allow us to choose a suitable modular variable for a particular kinematic region, where the integrals can be efficiently evaluated numerically using $\bar{q}$-expansion. This can be applied to accelerate the numerical evaluations of eMPLs~\cite{Walden:2020odh} and \emph{Lauricella functions}~\cite{Bezuglov:2023owj,Bera:2024hlq,Bezuglov:2025xol,Bezuglov:2025msm} of the elliptic type, which will be useful for phenomenological and mathematical purposes.

Our results provide a novel perspective on understanding canonical integrals and symbol letters for elliptic Feynman integrals. In particular, we find that both polylogarithmic and elliptic cases can be described in a unified language of pure functions. We expect that our results can be useful for symbol bootstrap in elliptic integral families~\cite{Morales:2022csr}. There are several directions worth future investigations opened up by this work. It would be interesting to investigate the relations between the construction presented here and the ones from Refs.~\cite{Adams:2018bsn,Pogel:2022vat,Gorges:2023zgv,Duhr:2025lbz,e-collaboration:2025frv,Bree:2025tug}. This will be given in a companion paper. An obvious next step is to extend the construction to multivariate cases involving elliptic sectors. For example, one can consider constructing canonical integrands with $\dlog \wedge \dlog$ and $\dlog \wedge \de$ forms for a bivariate family involving both polylogarithmic and elliptic sectors. Another possibility is to extend the construction to higher-genus cases, with pure polylogarithms on hyperelliptic curves~\cite{DHoker:2023vax,Baune:2024biq,DHoker:2024ozn,Baune:2024ber,DHoker:2025szl}. In this work, the linear combinations of $\de$-forms corresponding to Feynman integrals have been found by observation. It would be interesting to explore if there are simple selection rules for linear combinations of pure integrands which can be projected to Feynman integrals. This will be helpful in extending our approach to more complicated cases.

\section*{Acknowledgements}

We thank Jiaqi Chen, Hjalte Frellesvig, Stefan Weinzierl, and Xiaofeng Xu for useful discussions and comments on the draft. This work was supported in part by the National Natural Science Foundation of China under Grant No. 12375097, 12535003, 12347103, and the Fundamental Research Funds for the Central Universities.

\bibliography{refs}

\begin{thebibliography}{93}%
\makeatletter
\providecommand \@ifxundefined [1]{%
 \@ifx{#1\undefined}
}%
\providecommand \@ifnum [1]{%
 \ifnum #1\expandafter \@firstoftwo
 \else \expandafter \@secondoftwo
 \fi
}%
\providecommand \@ifx [1]{%
 \ifx #1\expandafter \@firstoftwo
 \else \expandafter \@secondoftwo
 \fi
}%
\providecommand \natexlab [1]{#1}%
\providecommand \enquote  [1]{``#1''}%
\providecommand \bibnamefont  [1]{#1}%
\providecommand \bibfnamefont [1]{#1}%
\providecommand \citenamefont [1]{#1}%
\providecommand \href@noop [0]{\@secondoftwo}%
\providecommand \href [0]{\begingroup \@sanitize@url \@href}%
\providecommand \@href[1]{\@@startlink{#1}\@@href}%
\providecommand \@@href[1]{\endgroup#1\@@endlink}%
\providecommand \@sanitize@url [0]{\catcode `\\12\catcode `\$12\catcode `\&12\catcode `\#12\catcode `\^12\catcode `\_12\catcode `\%12\relax}%
\providecommand \@@startlink[1]{}%
\providecommand \@@endlink[0]{}%
\providecommand \url  [0]{\begingroup\@sanitize@url \@url }%
\providecommand \@url [1]{\endgroup\@href {#1}{\urlprefix }}%
\providecommand \urlprefix  [0]{URL }%
\providecommand \Eprint [0]{\href }%
\providecommand \doibase [0]{https://doi.org/}%
\providecommand \selectlanguage [0]{\@gobble}%
\providecommand \bibinfo  [0]{\@secondoftwo}%
\providecommand \bibfield  [0]{\@secondoftwo}%
\providecommand \translation [1]{[#1]}%
\providecommand \BibitemOpen [0]{}%
\providecommand \bibitemStop [0]{}%
\providecommand \bibitemNoStop [0]{.\EOS\space}%
\providecommand \EOS [0]{\spacefactor3000\relax}%
\providecommand \BibitemShut  [1]{\csname bibitem#1\endcsname}%
\let\auto@bib@innerbib\@empty
\bibitem [{\citenamefont {Tkachov}(1981)}]{Tkachov:1981wb}%
  \BibitemOpen
  \bibfield  {author} {\bibinfo {author} {\bibfnamefont {F.~V.}\ \bibnamefont {Tkachov}},\ }\bibfield  {title} {\bibinfo {title} {{A theorem on analytical calculability of 4-loop renormalization group functions}},\ }\href {https://doi.org/10.1016/0370-2693(81)90288-4} {\bibfield  {journal} {\bibinfo  {journal} {Phys. Lett. B}\ }\textbf {\bibinfo {volume} {100}},\ \bibinfo {pages} {65} (\bibinfo {year} {1981})}\BibitemShut {NoStop}%
\bibitem [{\citenamefont {Chetyrkin}\ and\ \citenamefont {Tkachov}(1981)}]{Chetyrkin:1981qh}%
  \BibitemOpen
  \bibfield  {author} {\bibinfo {author} {\bibfnamefont {K.~G.}\ \bibnamefont {Chetyrkin}}\ and\ \bibinfo {author} {\bibfnamefont {F.~V.}\ \bibnamefont {Tkachov}},\ }\bibfield  {title} {\bibinfo {title} {{Integration by parts: The algorithm to calculate $\beta$-functions in 4 loops}},\ }\href {https://doi.org/10.1016/0550-3213(81)90199-1} {\bibfield  {journal} {\bibinfo  {journal} {Nucl. Phys. B}\ }\textbf {\bibinfo {volume} {192}},\ \bibinfo {pages} {159} (\bibinfo {year} {1981})}\BibitemShut {NoStop}%
\bibitem [{\citenamefont {Laporta}(2000)}]{Laporta:2000dsw}%
  \BibitemOpen
  \bibfield  {author} {\bibinfo {author} {\bibfnamefont {S.}~\bibnamefont {Laporta}},\ }\bibfield  {title} {\bibinfo {title} {{High-precision calculation of multiloop Feynman integrals by difference equations}},\ }\href {https://doi.org/10.1142/S0217751X00002159} {\bibfield  {journal} {\bibinfo  {journal} {Int. J. Mod. Phys. A}\ }\textbf {\bibinfo {volume} {15}},\ \bibinfo {pages} {5087} (\bibinfo {year} {2000})},\ \Eprint {https://arxiv.org/abs/hep-ph/0102033} {arXiv:hep-ph/0102033} \BibitemShut {NoStop}%
\bibitem [{\citenamefont {Kotikov}(1991{\natexlab{a}})}]{Kotikov:1990kg}%
  \BibitemOpen
  \bibfield  {author} {\bibinfo {author} {\bibfnamefont {A.~V.}\ \bibnamefont {Kotikov}},\ }\bibfield  {title} {\bibinfo {title} {{Differential equations method: New technique for massive Feynman diagrams calculation}},\ }\href {https://doi.org/10.1016/0370-2693(91)90413-K} {\bibfield  {journal} {\bibinfo  {journal} {Phys. Lett. B}\ }\textbf {\bibinfo {volume} {254}},\ \bibinfo {pages} {158} (\bibinfo {year} {1991}{\natexlab{a}})}\BibitemShut {NoStop}%
\bibitem [{\citenamefont {Kotikov}(1991{\natexlab{b}})}]{Kotikov:1991pm}%
  \BibitemOpen
  \bibfield  {author} {\bibinfo {author} {\bibfnamefont {A.~V.}\ \bibnamefont {Kotikov}},\ }\bibfield  {title} {\bibinfo {title} {{Differential equation method: The Calculation of N point Feynman diagrams}},\ }\href {https://doi.org/10.1016/0370-2693(91)90536-Y} {\bibfield  {journal} {\bibinfo  {journal} {Phys. Lett. B}\ }\textbf {\bibinfo {volume} {267}},\ \bibinfo {pages} {123} (\bibinfo {year} {1991}{\natexlab{b}})},\ \bibinfo {note} {[Erratum: Phys.Lett.B 295, 409--409 (1992)]}\BibitemShut {NoStop}%
\bibitem [{\citenamefont {Gehrmann}\ and\ \citenamefont {Remiddi}(2000)}]{Gehrmann:1999as}%
  \BibitemOpen
  \bibfield  {author} {\bibinfo {author} {\bibfnamefont {T.}~\bibnamefont {Gehrmann}}\ and\ \bibinfo {author} {\bibfnamefont {E.}~\bibnamefont {Remiddi}},\ }\bibfield  {title} {\bibinfo {title} {{Differential equations for two-loop four-point functions}},\ }\href {https://doi.org/10.1016/S0550-3213(00)00223-6} {\bibfield  {journal} {\bibinfo  {journal} {Nucl. Phys. B}\ }\textbf {\bibinfo {volume} {580}},\ \bibinfo {pages} {485} (\bibinfo {year} {2000})},\ \Eprint {https://arxiv.org/abs/hep-ph/9912329} {arXiv:hep-ph/9912329} \BibitemShut {NoStop}%
\bibitem [{\citenamefont {Henn}(2013)}]{Henn:2013pwa}%
  \BibitemOpen
  \bibfield  {author} {\bibinfo {author} {\bibfnamefont {J.~M.}\ \bibnamefont {Henn}},\ }\bibfield  {title} {\bibinfo {title} {{Multiloop integrals in dimensional regularization made simple}},\ }\href {https://doi.org/10.1103/PhysRevLett.110.251601} {\bibfield  {journal} {\bibinfo  {journal} {Phys. Rev. Lett.}\ }\textbf {\bibinfo {volume} {110}},\ \bibinfo {pages} {251601} (\bibinfo {year} {2013})},\ \Eprint {https://arxiv.org/abs/1304.1806} {arXiv:1304.1806 [hep-th]} \BibitemShut {NoStop}%
\bibitem [{\citenamefont {Chen}(1977)}]{Chen:1977oja}%
  \BibitemOpen
  \bibfield  {author} {\bibinfo {author} {\bibfnamefont {K.-T.}\ \bibnamefont {Chen}},\ }\bibfield  {title} {\bibinfo {title} {{Iterated path integrals}},\ }\href {https://doi.org/10.1090/S0002-9904-1977-14320-6} {\bibfield  {journal} {\bibinfo  {journal} {Bull. Am. Math. Soc.}\ }\textbf {\bibinfo {volume} {83}},\ \bibinfo {pages} {831} (\bibinfo {year} {1977})}\BibitemShut {NoStop}%
\bibitem [{\citenamefont {Goncharov}(1998)}]{Goncharov:1998kja}%
  \BibitemOpen
  \bibfield  {author} {\bibinfo {author} {\bibfnamefont {A.~B.}\ \bibnamefont {Goncharov}},\ }\bibfield  {title} {\bibinfo {title} {{Multiple polylogarithms, cyclotomy and modular complexes}},\ }\href {https://doi.org/10.4310/MRL.1998.v5.n4.a7} {\bibfield  {journal} {\bibinfo  {journal} {Math. Res. Lett.}\ }\textbf {\bibinfo {volume} {5}},\ \bibinfo {pages} {497} (\bibinfo {year} {1998})},\ \Eprint {https://arxiv.org/abs/1105.2076} {arXiv:1105.2076 [math.AG]} \BibitemShut {NoStop}%
\bibitem [{\citenamefont {Goncharov}(2001)}]{Goncharov:2001iea}%
  \BibitemOpen
  \bibfield  {author} {\bibinfo {author} {\bibfnamefont {A.~B.}\ \bibnamefont {Goncharov}},\ }\bibfield  {title} {\bibinfo {title} {{Multiple polylogarithms and mixed Tate motives}},\ }\href@noop {} {\  (\bibinfo {year} {2001})},\ \Eprint {https://arxiv.org/abs/math/0103059} {arXiv:math/0103059} \BibitemShut {NoStop}%
\bibitem [{\citenamefont {Lee}(2015)}]{Lee:2014ioa}%
  \BibitemOpen
  \bibfield  {author} {\bibinfo {author} {\bibfnamefont {R.~N.}\ \bibnamefont {Lee}},\ }\bibfield  {title} {\bibinfo {title} {{Reducing differential equations for multiloop master integrals}},\ }\href {https://doi.org/10.1007/JHEP04(2015)108} {\bibfield  {journal} {\bibinfo  {journal} {JHEP}\ }\textbf {\bibinfo {volume} {04}},\ \bibinfo {pages} {108}},\ \Eprint {https://arxiv.org/abs/1411.0911} {arXiv:1411.0911 [hep-ph]} \BibitemShut {NoStop}%
\bibitem [{\citenamefont {Lee}\ and\ \citenamefont {Pomeransky}(2017)}]{Lee:2017oca}%
  \BibitemOpen
  \bibfield  {author} {\bibinfo {author} {\bibfnamefont {R.~N.}\ \bibnamefont {Lee}}\ and\ \bibinfo {author} {\bibfnamefont {A.~A.}\ \bibnamefont {Pomeransky}},\ }\bibfield  {title} {\bibinfo {title} {{Normalized Fuchsian form on Riemann sphere and differential equations for multiloop integrals}},\ }\href@noop {} {\  (\bibinfo {year} {2017})},\ \Eprint {https://arxiv.org/abs/1707.07856} {arXiv:1707.07856 [hep-th]} \BibitemShut {NoStop}%
\bibitem [{\citenamefont {Gituliar}\ and\ \citenamefont {Magerya}(2017)}]{Gituliar:2017vzm}%
  \BibitemOpen
  \bibfield  {author} {\bibinfo {author} {\bibfnamefont {O.}~\bibnamefont {Gituliar}}\ and\ \bibinfo {author} {\bibfnamefont {V.}~\bibnamefont {Magerya}},\ }\bibfield  {title} {\bibinfo {title} {{Fuchsia: a tool for reducing differential equations for Feynman master integrals to epsilon form}},\ }\href {https://doi.org/10.1016/j.cpc.2017.05.004} {\bibfield  {journal} {\bibinfo  {journal} {Comput. Phys. Commun.}\ }\textbf {\bibinfo {volume} {219}},\ \bibinfo {pages} {329} (\bibinfo {year} {2017})},\ \Eprint {https://arxiv.org/abs/1701.04269} {arXiv:1701.04269 [hep-ph]} \BibitemShut {NoStop}%
\bibitem [{\citenamefont {Prausa}(2017)}]{Prausa:2017ltv}%
  \BibitemOpen
  \bibfield  {author} {\bibinfo {author} {\bibfnamefont {M.}~\bibnamefont {Prausa}},\ }\bibfield  {title} {\bibinfo {title} {{epsilon: A tool to find a canonical basis of master integrals}},\ }\href {https://doi.org/10.1016/j.cpc.2017.05.026} {\bibfield  {journal} {\bibinfo  {journal} {Comput. Phys. Commun.}\ }\textbf {\bibinfo {volume} {219}},\ \bibinfo {pages} {361} (\bibinfo {year} {2017})},\ \Eprint {https://arxiv.org/abs/1701.00725} {arXiv:1701.00725 [hep-ph]} \BibitemShut {NoStop}%
\bibitem [{\citenamefont {Meyer}(2018)}]{Meyer:2017joq}%
  \BibitemOpen
  \bibfield  {author} {\bibinfo {author} {\bibfnamefont {C.}~\bibnamefont {Meyer}},\ }\bibfield  {title} {\bibinfo {title} {{Algorithmic transformation of multi-loop master integrals to a canonical basis with CANONICA}},\ }\href {https://doi.org/10.1016/j.cpc.2017.09.014} {\bibfield  {journal} {\bibinfo  {journal} {Comput. Phys. Commun.}\ }\textbf {\bibinfo {volume} {222}},\ \bibinfo {pages} {295} (\bibinfo {year} {2018})},\ \Eprint {https://arxiv.org/abs/1705.06252} {arXiv:1705.06252 [hep-ph]} \BibitemShut {NoStop}%
\bibitem [{\citenamefont {Lee}(2021)}]{Lee:2020zfb}%
  \BibitemOpen
  \bibfield  {author} {\bibinfo {author} {\bibfnamefont {R.~N.}\ \bibnamefont {Lee}},\ }\bibfield  {title} {\bibinfo {title} {{Libra: A package for transformation of differential systems for multiloop integrals}},\ }\href {https://doi.org/10.1016/j.cpc.2021.108058} {\bibfield  {journal} {\bibinfo  {journal} {Comput. Phys. Commun.}\ }\textbf {\bibinfo {volume} {267}},\ \bibinfo {pages} {108058} (\bibinfo {year} {2021})},\ \Eprint {https://arxiv.org/abs/2012.00279} {arXiv:2012.00279 [hep-ph]} \BibitemShut {NoStop}%
\bibitem [{\citenamefont {Dlapa}\ \emph {et~al.}(2020)\citenamefont {Dlapa}, \citenamefont {Henn},\ and\ \citenamefont {Yan}}]{Dlapa:2020cwj}%
  \BibitemOpen
  \bibfield  {author} {\bibinfo {author} {\bibfnamefont {C.}~\bibnamefont {Dlapa}}, \bibinfo {author} {\bibfnamefont {J.}~\bibnamefont {Henn}},\ and\ \bibinfo {author} {\bibfnamefont {K.}~\bibnamefont {Yan}},\ }\bibfield  {title} {\bibinfo {title} {{Deriving canonical differential equations for Feynman integrals from a single uniform weight integral}},\ }\href {https://doi.org/10.1007/JHEP05(2020)025} {\bibfield  {journal} {\bibinfo  {journal} {JHEP}\ }\textbf {\bibinfo {volume} {05}},\ \bibinfo {pages} {025}},\ \Eprint {https://arxiv.org/abs/2002.02340} {arXiv:2002.02340 [hep-ph]} \BibitemShut {NoStop}%
\bibitem [{\citenamefont {Henn}\ \emph {et~al.}(2020)\citenamefont {Henn}, \citenamefont {Mistlberger}, \citenamefont {Smirnov},\ and\ \citenamefont {Wasser}}]{Henn:2020lye}%
  \BibitemOpen
  \bibfield  {author} {\bibinfo {author} {\bibfnamefont {J.}~\bibnamefont {Henn}}, \bibinfo {author} {\bibfnamefont {B.}~\bibnamefont {Mistlberger}}, \bibinfo {author} {\bibfnamefont {V.~A.}\ \bibnamefont {Smirnov}},\ and\ \bibinfo {author} {\bibfnamefont {P.}~\bibnamefont {Wasser}},\ }\bibfield  {title} {\bibinfo {title} {{Constructing d-log integrands and computing master integrals for three-loop four-particle scattering}},\ }\href {https://doi.org/10.1007/JHEP04(2020)167} {\bibfield  {journal} {\bibinfo  {journal} {JHEP}\ }\textbf {\bibinfo {volume} {04}},\ \bibinfo {pages} {167}},\ \Eprint {https://arxiv.org/abs/2002.09492} {arXiv:2002.09492 [hep-ph]} \BibitemShut {NoStop}%
\bibitem [{\citenamefont {Chen}\ \emph {et~al.}(2021)\citenamefont {Chen}, \citenamefont {Jiang}, \citenamefont {Xu},\ and\ \citenamefont {Yang}}]{Chen:2020uyk}%
  \BibitemOpen
  \bibfield  {author} {\bibinfo {author} {\bibfnamefont {J.}~\bibnamefont {Chen}}, \bibinfo {author} {\bibfnamefont {X.}~\bibnamefont {Jiang}}, \bibinfo {author} {\bibfnamefont {X.}~\bibnamefont {Xu}},\ and\ \bibinfo {author} {\bibfnamefont {L.~L.}\ \bibnamefont {Yang}},\ }\bibfield  {title} {\bibinfo {title} {{Constructing canonical Feynman integrals with intersection theory}},\ }\href {https://doi.org/10.1016/j.physletb.2021.136085} {\bibfield  {journal} {\bibinfo  {journal} {Phys. Lett. B}\ }\textbf {\bibinfo {volume} {814}},\ \bibinfo {pages} {136085} (\bibinfo {year} {2021})},\ \Eprint {https://arxiv.org/abs/2008.03045} {arXiv:2008.03045 [hep-th]} \BibitemShut {NoStop}%
\bibitem [{\citenamefont {Chen}\ \emph {et~al.}(2022)\citenamefont {Chen}, \citenamefont {Jiang}, \citenamefont {Ma}, \citenamefont {Xu},\ and\ \citenamefont {Yang}}]{Chen:2022lzr}%
  \BibitemOpen
  \bibfield  {author} {\bibinfo {author} {\bibfnamefont {J.}~\bibnamefont {Chen}}, \bibinfo {author} {\bibfnamefont {X.}~\bibnamefont {Jiang}}, \bibinfo {author} {\bibfnamefont {C.}~\bibnamefont {Ma}}, \bibinfo {author} {\bibfnamefont {X.}~\bibnamefont {Xu}},\ and\ \bibinfo {author} {\bibfnamefont {L.~L.}\ \bibnamefont {Yang}},\ }\bibfield  {title} {\bibinfo {title} {{Baikov representations, intersection theory, and canonical Feynman integrals}},\ }\href {https://doi.org/10.1007/JHEP07(2022)066} {\bibfield  {journal} {\bibinfo  {journal} {JHEP}\ }\textbf {\bibinfo {volume} {07}},\ \bibinfo {pages} {066}},\ \Eprint {https://arxiv.org/abs/2202.08127} {arXiv:2202.08127 [hep-th]} \BibitemShut {NoStop}%
\bibitem [{\citenamefont {Cachazo}(2008)}]{Cachazo:2008vp}%
  \BibitemOpen
  \bibfield  {author} {\bibinfo {author} {\bibfnamefont {F.}~\bibnamefont {Cachazo}},\ }\bibfield  {title} {\bibinfo {title} {{Sharpening The Leading Singularity}},\ }\href@noop {} {\  (\bibinfo {year} {2008})},\ \Eprint {https://arxiv.org/abs/0803.1988} {arXiv:0803.1988 [hep-th]} \BibitemShut {NoStop}%
\bibitem [{\citenamefont {Arkani-Hamed}\ \emph {et~al.}(2012)\citenamefont {Arkani-Hamed}, \citenamefont {Bourjaily}, \citenamefont {Cachazo},\ and\ \citenamefont {Trnka}}]{Arkani-Hamed:2010pyv}%
  \BibitemOpen
  \bibfield  {author} {\bibinfo {author} {\bibfnamefont {N.}~\bibnamefont {Arkani-Hamed}}, \bibinfo {author} {\bibfnamefont {J.~L.}\ \bibnamefont {Bourjaily}}, \bibinfo {author} {\bibfnamefont {F.}~\bibnamefont {Cachazo}},\ and\ \bibinfo {author} {\bibfnamefont {J.}~\bibnamefont {Trnka}},\ }\bibfield  {title} {\bibinfo {title} {{Local Integrals for Planar Scattering Amplitudes}},\ }\href {https://doi.org/10.1007/JHEP06(2012)125} {\bibfield  {journal} {\bibinfo  {journal} {JHEP}\ }\textbf {\bibinfo {volume} {06}},\ \bibinfo {pages} {125}},\ \Eprint {https://arxiv.org/abs/1012.6032} {arXiv:1012.6032 [hep-th]} \BibitemShut {NoStop}%
\bibitem [{\citenamefont {Primo}\ and\ \citenamefont {Tancredi}(2017{\natexlab{a}})}]{Primo:2016ebd}%
  \BibitemOpen
  \bibfield  {author} {\bibinfo {author} {\bibfnamefont {A.}~\bibnamefont {Primo}}\ and\ \bibinfo {author} {\bibfnamefont {L.}~\bibnamefont {Tancredi}},\ }\bibfield  {title} {\bibinfo {title} {{On the maximal cut of Feynman integrals and the solution of their differential equations}},\ }\href {https://doi.org/10.1016/j.nuclphysb.2016.12.021} {\bibfield  {journal} {\bibinfo  {journal} {Nucl. Phys. B}\ }\textbf {\bibinfo {volume} {916}},\ \bibinfo {pages} {94} (\bibinfo {year} {2017}{\natexlab{a}})},\ \Eprint {https://arxiv.org/abs/1610.08397} {arXiv:1610.08397 [hep-ph]} \BibitemShut {NoStop}%
\bibitem [{\citenamefont {Frellesvig}\ and\ \citenamefont {Papadopoulos}(2017)}]{Frellesvig:2017aai}%
  \BibitemOpen
  \bibfield  {author} {\bibinfo {author} {\bibfnamefont {H.}~\bibnamefont {Frellesvig}}\ and\ \bibinfo {author} {\bibfnamefont {C.~G.}\ \bibnamefont {Papadopoulos}},\ }\bibfield  {title} {\bibinfo {title} {{Cuts of Feynman Integrals in Baikov representation}},\ }\href {https://doi.org/10.1007/JHEP04(2017)083} {\bibfield  {journal} {\bibinfo  {journal} {JHEP}\ }\textbf {\bibinfo {volume} {04}},\ \bibinfo {pages} {083}},\ \Eprint {https://arxiv.org/abs/1701.07356} {arXiv:1701.07356 [hep-ph]} \BibitemShut {NoStop}%
\bibitem [{\citenamefont {Bosma}\ \emph {et~al.}(2017)\citenamefont {Bosma}, \citenamefont {Sogaard},\ and\ \citenamefont {Zhang}}]{Bosma:2017ens}%
  \BibitemOpen
  \bibfield  {author} {\bibinfo {author} {\bibfnamefont {J.}~\bibnamefont {Bosma}}, \bibinfo {author} {\bibfnamefont {M.}~\bibnamefont {Sogaard}},\ and\ \bibinfo {author} {\bibfnamefont {Y.}~\bibnamefont {Zhang}},\ }\bibfield  {title} {\bibinfo {title} {{Maximal Cuts in Arbitrary Dimension}},\ }\href {https://doi.org/10.1007/JHEP08(2017)051} {\bibfield  {journal} {\bibinfo  {journal} {JHEP}\ }\textbf {\bibinfo {volume} {08}},\ \bibinfo {pages} {051}},\ \Eprint {https://arxiv.org/abs/1704.04255} {arXiv:1704.04255 [hep-th]} \BibitemShut {NoStop}%
\bibitem [{\citenamefont {Primo}\ and\ \citenamefont {Tancredi}(2017{\natexlab{b}})}]{Primo:2017ipr}%
  \BibitemOpen
  \bibfield  {author} {\bibinfo {author} {\bibfnamefont {A.}~\bibnamefont {Primo}}\ and\ \bibinfo {author} {\bibfnamefont {L.}~\bibnamefont {Tancredi}},\ }\bibfield  {title} {\bibinfo {title} {{Maximal cuts and differential equations for Feynman integrals. An application to the three-loop massive banana graph}},\ }\href {https://doi.org/10.1016/j.nuclphysb.2017.05.018} {\bibfield  {journal} {\bibinfo  {journal} {Nucl. Phys. B}\ }\textbf {\bibinfo {volume} {921}},\ \bibinfo {pages} {316} (\bibinfo {year} {2017}{\natexlab{b}})},\ \Eprint {https://arxiv.org/abs/1704.05465} {arXiv:1704.05465 [hep-ph]} \BibitemShut {NoStop}%
\bibitem [{\citenamefont {Harley}\ \emph {et~al.}(2017)\citenamefont {Harley}, \citenamefont {Moriello},\ and\ \citenamefont {Schabinger}}]{Harley:2017qut}%
  \BibitemOpen
  \bibfield  {author} {\bibinfo {author} {\bibfnamefont {M.}~\bibnamefont {Harley}}, \bibinfo {author} {\bibfnamefont {F.}~\bibnamefont {Moriello}},\ and\ \bibinfo {author} {\bibfnamefont {R.~M.}\ \bibnamefont {Schabinger}},\ }\bibfield  {title} {\bibinfo {title} {{Baikov-Lee Representations Of Cut Feynman Integrals}},\ }\href {https://doi.org/10.1007/JHEP06(2017)049} {\bibfield  {journal} {\bibinfo  {journal} {JHEP}\ }\textbf {\bibinfo {volume} {06}},\ \bibinfo {pages} {049}},\ \Eprint {https://arxiv.org/abs/1705.03478} {arXiv:1705.03478 [hep-ph]} \BibitemShut {NoStop}%
\bibitem [{\citenamefont {Baikov}(1996)}]{Baikov:1996rk}%
  \BibitemOpen
  \bibfield  {author} {\bibinfo {author} {\bibfnamefont {P.~A.}\ \bibnamefont {Baikov}},\ }\bibfield  {title} {\bibinfo {title} {{Explicit solutions of the three loop vacuum integral recurrence relations}},\ }\href {https://doi.org/10.1016/0370-2693(96)00835-0} {\bibfield  {journal} {\bibinfo  {journal} {Phys. Lett. B}\ }\textbf {\bibinfo {volume} {385}},\ \bibinfo {pages} {404} (\bibinfo {year} {1996})},\ \Eprint {https://arxiv.org/abs/hep-ph/9603267} {arXiv:hep-ph/9603267} \BibitemShut {NoStop}%
\bibitem [{\citenamefont {Baikov}(1997)}]{Baikov:1996iu}%
  \BibitemOpen
  \bibfield  {author} {\bibinfo {author} {\bibfnamefont {P.~A.}\ \bibnamefont {Baikov}},\ }\bibfield  {title} {\bibinfo {title} {{Explicit solutions of the multiloop integral recurrence relations and its application}},\ }\href {https://doi.org/10.1016/S0168-9002(97)00126-5} {\bibfield  {journal} {\bibinfo  {journal} {Nucl. Instrum. Meth. A}\ }\textbf {\bibinfo {volume} {389}},\ \bibinfo {pages} {347} (\bibinfo {year} {1997})},\ \Eprint {https://arxiv.org/abs/hep-ph/9611449} {arXiv:hep-ph/9611449} \BibitemShut {NoStop}%
\bibitem [{\citenamefont {Mastrolia}\ and\ \citenamefont {Mizera}(2019)}]{Mastrolia:2018uzb}%
  \BibitemOpen
  \bibfield  {author} {\bibinfo {author} {\bibfnamefont {P.}~\bibnamefont {Mastrolia}}\ and\ \bibinfo {author} {\bibfnamefont {S.}~\bibnamefont {Mizera}},\ }\bibfield  {title} {\bibinfo {title} {{Feynman Integrals and Intersection Theory}},\ }\href {https://doi.org/10.1007/JHEP02(2019)139} {\bibfield  {journal} {\bibinfo  {journal} {JHEP}\ }\textbf {\bibinfo {volume} {02}},\ \bibinfo {pages} {139}},\ \Eprint {https://arxiv.org/abs/1810.03818} {arXiv:1810.03818 [hep-th]} \BibitemShut {NoStop}%
\bibitem [{\citenamefont {Frellesvig}\ \emph {et~al.}(2019)\citenamefont {Frellesvig}, \citenamefont {Gasparotto}, \citenamefont {Mandal}, \citenamefont {Mastrolia}, \citenamefont {Mattiazzi},\ and\ \citenamefont {Mizera}}]{Frellesvig:2019uqt}%
  \BibitemOpen
  \bibfield  {author} {\bibinfo {author} {\bibfnamefont {H.}~\bibnamefont {Frellesvig}}, \bibinfo {author} {\bibfnamefont {F.}~\bibnamefont {Gasparotto}}, \bibinfo {author} {\bibfnamefont {M.~K.}\ \bibnamefont {Mandal}}, \bibinfo {author} {\bibfnamefont {P.}~\bibnamefont {Mastrolia}}, \bibinfo {author} {\bibfnamefont {L.}~\bibnamefont {Mattiazzi}},\ and\ \bibinfo {author} {\bibfnamefont {S.}~\bibnamefont {Mizera}},\ }\bibfield  {title} {\bibinfo {title} {{Vector Space of Feynman Integrals and Multivariate Intersection Numbers}},\ }\href {https://doi.org/10.1103/PhysRevLett.123.201602} {\bibfield  {journal} {\bibinfo  {journal} {Phys. Rev. Lett.}\ }\textbf {\bibinfo {volume} {123}},\ \bibinfo {pages} {201602} (\bibinfo {year} {2019})},\ \Eprint {https://arxiv.org/abs/1907.02000} {arXiv:1907.02000 [hep-th]} \BibitemShut {NoStop}%
\bibitem [{\citenamefont {Bourjaily}\ \emph {et~al.}(2022)\citenamefont {Bourjaily} \emph {et~al.}}]{Bourjaily:2022bwx}%
  \BibitemOpen
  \bibfield  {author} {\bibinfo {author} {\bibfnamefont {J.~L.}\ \bibnamefont {Bourjaily}} \emph {et~al.},\ }\bibfield  {title} {\bibinfo {title} {{Functions Beyond Multiple Polylogarithms for Precision Collider Physics}},\ }in\ \href@noop {} {\emph {\bibinfo {booktitle} {{Snowmass 2021}}}}\ (\bibinfo {year} {2022})\ \Eprint {https://arxiv.org/abs/2203.07088} {arXiv:2203.07088 [hep-ph]} \BibitemShut {NoStop}%
\bibitem [{\citenamefont {Adams}\ and\ \citenamefont {Weinzierl}(2018{\natexlab{a}})}]{Adams:2018yfj}%
  \BibitemOpen
  \bibfield  {author} {\bibinfo {author} {\bibfnamefont {L.}~\bibnamefont {Adams}}\ and\ \bibinfo {author} {\bibfnamefont {S.}~\bibnamefont {Weinzierl}},\ }\bibfield  {title} {\bibinfo {title} {{The $\varepsilon$-form of the differential equations for Feynman integrals in the elliptic case}},\ }\href {https://doi.org/10.1016/j.physletb.2018.04.002} {\bibfield  {journal} {\bibinfo  {journal} {Phys. Lett. B}\ }\textbf {\bibinfo {volume} {781}},\ \bibinfo {pages} {270} (\bibinfo {year} {2018}{\natexlab{a}})},\ \Eprint {https://arxiv.org/abs/1802.05020} {arXiv:1802.05020 [hep-ph]} \BibitemShut {NoStop}%
\bibitem [{\citenamefont {Adams}\ \emph {et~al.}(2018)\citenamefont {Adams}, \citenamefont {Chaubey},\ and\ \citenamefont {Weinzierl}}]{Adams:2018bsn}%
  \BibitemOpen
  \bibfield  {author} {\bibinfo {author} {\bibfnamefont {L.}~\bibnamefont {Adams}}, \bibinfo {author} {\bibfnamefont {E.}~\bibnamefont {Chaubey}},\ and\ \bibinfo {author} {\bibfnamefont {S.}~\bibnamefont {Weinzierl}},\ }\bibfield  {title} {\bibinfo {title} {{Planar Double Box Integral for Top Pair Production with a Closed Top Loop to all orders in the Dimensional Regularization Parameter}},\ }\href {https://doi.org/10.1103/PhysRevLett.121.142001} {\bibfield  {journal} {\bibinfo  {journal} {Phys. Rev. Lett.}\ }\textbf {\bibinfo {volume} {121}},\ \bibinfo {pages} {142001} (\bibinfo {year} {2018})},\ \Eprint {https://arxiv.org/abs/1804.11144} {arXiv:1804.11144 [hep-ph]} \BibitemShut {NoStop}%
\bibitem [{\citenamefont {Giroux}\ and\ \citenamefont {Pokraka}(2023)}]{Giroux:2022wav}%
  \BibitemOpen
  \bibfield  {author} {\bibinfo {author} {\bibfnamefont {M.}~\bibnamefont {Giroux}}\ and\ \bibinfo {author} {\bibfnamefont {A.}~\bibnamefont {Pokraka}},\ }\bibfield  {title} {\bibinfo {title} {{Loop-by-loop differential equations for dual (elliptic) Feynman integrals}},\ }\href {https://doi.org/10.1007/JHEP03(2023)155} {\bibfield  {journal} {\bibinfo  {journal} {JHEP}\ }\textbf {\bibinfo {volume} {03}},\ \bibinfo {pages} {155}},\ \Eprint {https://arxiv.org/abs/2210.09898} {arXiv:2210.09898 [hep-th]} \BibitemShut {NoStop}%
\bibitem [{\citenamefont {Dlapa}\ \emph {et~al.}(2023)\citenamefont {Dlapa}, \citenamefont {Henn},\ and\ \citenamefont {Wagner}}]{Dlapa:2022wdu}%
  \BibitemOpen
  \bibfield  {author} {\bibinfo {author} {\bibfnamefont {C.}~\bibnamefont {Dlapa}}, \bibinfo {author} {\bibfnamefont {J.~M.}\ \bibnamefont {Henn}},\ and\ \bibinfo {author} {\bibfnamefont {F.~J.}\ \bibnamefont {Wagner}},\ }\bibfield  {title} {\bibinfo {title} {{An algorithmic approach to finding canonical differential equations for elliptic Feynman integrals}},\ }\href {https://doi.org/10.1007/JHEP08(2023)120} {\bibfield  {journal} {\bibinfo  {journal} {JHEP}\ }\textbf {\bibinfo {volume} {08}},\ \bibinfo {pages} {120}},\ \Eprint {https://arxiv.org/abs/2211.16357} {arXiv:2211.16357 [hep-ph]} \BibitemShut {NoStop}%
\bibitem [{\citenamefont {P\"ogel}\ \emph {et~al.}(2023{\natexlab{a}})\citenamefont {P\"ogel}, \citenamefont {Wang},\ and\ \citenamefont {Weinzierl}}]{Pogel:2022vat}%
  \BibitemOpen
  \bibfield  {author} {\bibinfo {author} {\bibfnamefont {S.}~\bibnamefont {P\"ogel}}, \bibinfo {author} {\bibfnamefont {X.}~\bibnamefont {Wang}},\ and\ \bibinfo {author} {\bibfnamefont {S.}~\bibnamefont {Weinzierl}},\ }\bibfield  {title} {\bibinfo {title} {{Bananas of equal mass: any loop, any order in the dimensional regularisation parameter}},\ }\href {https://doi.org/10.1007/JHEP04(2023)117} {\bibfield  {journal} {\bibinfo  {journal} {JHEP}\ }\textbf {\bibinfo {volume} {04}},\ \bibinfo {pages} {117}},\ \Eprint {https://arxiv.org/abs/2212.08908} {arXiv:2212.08908 [hep-th]} \BibitemShut {NoStop}%
\bibitem [{\citenamefont {G\"orges}\ \emph {et~al.}(2023)\citenamefont {G\"orges}, \citenamefont {Nega}, \citenamefont {Tancredi},\ and\ \citenamefont {Wagner}}]{Gorges:2023zgv}%
  \BibitemOpen
  \bibfield  {author} {\bibinfo {author} {\bibfnamefont {L.}~\bibnamefont {G\"orges}}, \bibinfo {author} {\bibfnamefont {C.}~\bibnamefont {Nega}}, \bibinfo {author} {\bibfnamefont {L.}~\bibnamefont {Tancredi}},\ and\ \bibinfo {author} {\bibfnamefont {F.~J.}\ \bibnamefont {Wagner}},\ }\bibfield  {title} {\bibinfo {title} {{On a procedure to derive \ensuremath{\epsilon}-factorised differential equations beyond polylogarithms}},\ }\href {https://doi.org/10.1007/JHEP07(2023)206} {\bibfield  {journal} {\bibinfo  {journal} {JHEP}\ }\textbf {\bibinfo {volume} {07}},\ \bibinfo {pages} {206}},\ \Eprint {https://arxiv.org/abs/2305.14090} {arXiv:2305.14090 [hep-th]} \BibitemShut {NoStop}%
\bibitem [{\citenamefont {Duhr}\ \emph {et~al.}(2025{\natexlab{a}})\citenamefont {Duhr}, \citenamefont {Maggio}, \citenamefont {Nega}, \citenamefont {Sauer}, \citenamefont {Tancredi},\ and\ \citenamefont {Wagner}}]{Duhr:2025lbz}%
  \BibitemOpen
  \bibfield  {author} {\bibinfo {author} {\bibfnamefont {C.}~\bibnamefont {Duhr}}, \bibinfo {author} {\bibfnamefont {S.}~\bibnamefont {Maggio}}, \bibinfo {author} {\bibfnamefont {C.}~\bibnamefont {Nega}}, \bibinfo {author} {\bibfnamefont {B.}~\bibnamefont {Sauer}}, \bibinfo {author} {\bibfnamefont {L.}~\bibnamefont {Tancredi}},\ and\ \bibinfo {author} {\bibfnamefont {F.~J.}\ \bibnamefont {Wagner}},\ }\bibfield  {title} {\bibinfo {title} {{Aspects of canonical differential equations for Calabi-Yau geometries and beyond}},\ }\href {https://doi.org/10.1007/JHEP06(2025)128} {\bibfield  {journal} {\bibinfo  {journal} {JHEP}\ }\textbf {\bibinfo {volume} {06}},\ \bibinfo {pages} {128}},\ \Eprint {https://arxiv.org/abs/2503.20655} {arXiv:2503.20655 [hep-th]} \BibitemShut {NoStop}%
\bibitem [{\citenamefont {Maggio}\ and\ \citenamefont {Sohnle}(2025)}]{Maggio:2025jel}%
  \BibitemOpen
  \bibfield  {author} {\bibinfo {author} {\bibfnamefont {S.}~\bibnamefont {Maggio}}\ and\ \bibinfo {author} {\bibfnamefont {Y.}~\bibnamefont {Sohnle}},\ }\bibfield  {title} {\bibinfo {title} {{On canonical differential equations for Calabi-Yau multi-scale Feynman integrals}},\ }\href {https://doi.org/10.1007/JHEP10(2025)202} {\bibfield  {journal} {\bibinfo  {journal} {JHEP}\ }\textbf {\bibinfo {volume} {10}},\ \bibinfo {pages} {202}},\ \Eprint {https://arxiv.org/abs/2504.17757} {arXiv:2504.17757 [hep-th]} \BibitemShut {NoStop}%
\bibitem [{\citenamefont {Bree}\ \emph {et~al.}(2025{\natexlab{a}})\citenamefont {Bree} \emph {et~al.}}]{e-collaboration:2025frv}%
  \BibitemOpen
  \bibfield  {author} {\bibinfo {author} {\bibfnamefont {I.}~\bibnamefont {Bree}} \emph {et~al.} (\bibinfo {collaboration} {{\ensuremath{\varepsilon}}-collaboration}),\ }\bibfield  {title} {\bibinfo {title} {{The geometric bookkeeping guide to Feynman integral reduction and $\varepsilon$-factorised differential equations}},\ }\href@noop {} {\  (\bibinfo {year} {2025}{\natexlab{a}})},\ \Eprint {https://arxiv.org/abs/2506.09124} {arXiv:2506.09124 [hep-th]} \BibitemShut {NoStop}%
\bibitem [{\citenamefont {Duhr}\ \emph {et~al.}(2025{\natexlab{b}})\citenamefont {Duhr}, \citenamefont {Maggio}, \citenamefont {Porkert}, \citenamefont {Semper}, \citenamefont {Sohnle},\ and\ \citenamefont {Stawinski}}]{Duhr:2025xyy}%
  \BibitemOpen
  \bibfield  {author} {\bibinfo {author} {\bibfnamefont {C.}~\bibnamefont {Duhr}}, \bibinfo {author} {\bibfnamefont {S.}~\bibnamefont {Maggio}}, \bibinfo {author} {\bibfnamefont {F.}~\bibnamefont {Porkert}}, \bibinfo {author} {\bibfnamefont {C.}~\bibnamefont {Semper}}, \bibinfo {author} {\bibfnamefont {Y.}~\bibnamefont {Sohnle}},\ and\ \bibinfo {author} {\bibfnamefont {S.~F.}\ \bibnamefont {Stawinski}},\ }\bibfield  {title} {\bibinfo {title} {{Canonical differential equations and intersection matrices}},\ }\href@noop {} {\  (\bibinfo {year} {2025}{\natexlab{b}})},\ \Eprint {https://arxiv.org/abs/2509.17787} {arXiv:2509.17787 [hep-th]} \BibitemShut {NoStop}%
\bibitem [{\citenamefont {Bree}\ \emph {et~al.}(2025{\natexlab{b}})\citenamefont {Bree} \emph {et~al.}}]{Bree:2025tug}%
  \BibitemOpen
  \bibfield  {author} {\bibinfo {author} {\bibfnamefont {I.}~\bibnamefont {Bree}} \emph {et~al.},\ }\bibfield  {title} {\bibinfo {title} {{New algorithms for Feynman integral reduction and $\varepsilon$-factorised differential equations}},\ }\href@noop {} {\  (\bibinfo {year} {2025}{\natexlab{b}})},\ \Eprint {https://arxiv.org/abs/2511.15381} {arXiv:2511.15381 [hep-th]} \BibitemShut {NoStop}%
\bibitem [{\citenamefont {Bogner}\ \emph {et~al.}(2020)\citenamefont {Bogner}, \citenamefont {M\"uller-Stach},\ and\ \citenamefont {Weinzierl}}]{Bogner:2019lfa}%
  \BibitemOpen
  \bibfield  {author} {\bibinfo {author} {\bibfnamefont {C.}~\bibnamefont {Bogner}}, \bibinfo {author} {\bibfnamefont {S.}~\bibnamefont {M\"uller-Stach}},\ and\ \bibinfo {author} {\bibfnamefont {S.}~\bibnamefont {Weinzierl}},\ }\bibfield  {title} {\bibinfo {title} {{The unequal mass sunrise integral expressed through iterated integrals on $\overline{\mathcal M}_{1,3}$}},\ }\href {https://doi.org/10.1016/j.nuclphysb.2020.114991} {\bibfield  {journal} {\bibinfo  {journal} {Nucl. Phys. B}\ }\textbf {\bibinfo {volume} {954}},\ \bibinfo {pages} {114991} (\bibinfo {year} {2020})},\ \Eprint {https://arxiv.org/abs/1907.01251} {arXiv:1907.01251 [hep-th]} \BibitemShut {NoStop}%
\bibitem [{\citenamefont {M\"uller}\ and\ \citenamefont {Weinzierl}(2022)}]{Muller:2022gec}%
  \BibitemOpen
  \bibfield  {author} {\bibinfo {author} {\bibfnamefont {H.}~\bibnamefont {M\"uller}}\ and\ \bibinfo {author} {\bibfnamefont {S.}~\bibnamefont {Weinzierl}},\ }\bibfield  {title} {\bibinfo {title} {{A Feynman integral depending on two elliptic curves}},\ }\href {https://doi.org/10.1007/JHEP07(2022)101} {\bibfield  {journal} {\bibinfo  {journal} {JHEP}\ }\textbf {\bibinfo {volume} {07}},\ \bibinfo {pages} {101}},\ \Eprint {https://arxiv.org/abs/2205.04818} {arXiv:2205.04818 [hep-th]} \BibitemShut {NoStop}%
\bibitem [{\citenamefont {P\"ogel}\ \emph {et~al.}(2022)\citenamefont {P\"ogel}, \citenamefont {Wang},\ and\ \citenamefont {Weinzierl}}]{Pogel:2022yat}%
  \BibitemOpen
  \bibfield  {author} {\bibinfo {author} {\bibfnamefont {S.}~\bibnamefont {P\"ogel}}, \bibinfo {author} {\bibfnamefont {X.}~\bibnamefont {Wang}},\ and\ \bibinfo {author} {\bibfnamefont {S.}~\bibnamefont {Weinzierl}},\ }\bibfield  {title} {\bibinfo {title} {{The three-loop equal-mass banana integral in \ensuremath{\varepsilon}-factorised form with meromorphic modular forms}},\ }\href {https://doi.org/10.1007/JHEP09(2022)062} {\bibfield  {journal} {\bibinfo  {journal} {JHEP}\ }\textbf {\bibinfo {volume} {09}},\ \bibinfo {pages} {062}},\ \Eprint {https://arxiv.org/abs/2207.12893} {arXiv:2207.12893 [hep-th]} \BibitemShut {NoStop}%
\bibitem [{\citenamefont {P\"ogel}\ \emph {et~al.}(2023{\natexlab{b}})\citenamefont {P\"ogel}, \citenamefont {Wang},\ and\ \citenamefont {Weinzierl}}]{Pogel:2022ken}%
  \BibitemOpen
  \bibfield  {author} {\bibinfo {author} {\bibfnamefont {S.}~\bibnamefont {P\"ogel}}, \bibinfo {author} {\bibfnamefont {X.}~\bibnamefont {Wang}},\ and\ \bibinfo {author} {\bibfnamefont {S.}~\bibnamefont {Weinzierl}},\ }\bibfield  {title} {\bibinfo {title} {{Taming Calabi-Yau Feynman Integrals: The Four-Loop Equal-Mass Banana Integral}},\ }\href {https://doi.org/10.1103/PhysRevLett.130.101601} {\bibfield  {journal} {\bibinfo  {journal} {Phys. Rev. Lett.}\ }\textbf {\bibinfo {volume} {130}},\ \bibinfo {pages} {101601} (\bibinfo {year} {2023}{\natexlab{b}})},\ \Eprint {https://arxiv.org/abs/2211.04292} {arXiv:2211.04292 [hep-th]} \BibitemShut {NoStop}%
\bibitem [{\citenamefont {Jiang}\ \emph {et~al.}(2023)\citenamefont {Jiang}, \citenamefont {Wang}, \citenamefont {Yang},\ and\ \citenamefont {Zhao}}]{Jiang:2023jmk}%
  \BibitemOpen
  \bibfield  {author} {\bibinfo {author} {\bibfnamefont {X.}~\bibnamefont {Jiang}}, \bibinfo {author} {\bibfnamefont {X.}~\bibnamefont {Wang}}, \bibinfo {author} {\bibfnamefont {L.~L.}\ \bibnamefont {Yang}},\ and\ \bibinfo {author} {\bibfnamefont {J.}~\bibnamefont {Zhao}},\ }\bibfield  {title} {\bibinfo {title} {{\ensuremath{\varepsilon}-factorized differential equations for two-loop non-planar triangle Feynman integrals with elliptic curves}},\ }\href {https://doi.org/10.1007/JHEP09(2023)187} {\bibfield  {journal} {\bibinfo  {journal} {JHEP}\ }\textbf {\bibinfo {volume} {09}},\ \bibinfo {pages} {187}},\ \Eprint {https://arxiv.org/abs/2305.13951} {arXiv:2305.13951 [hep-th]} \BibitemShut {NoStop}%
\bibitem [{\citenamefont {Delto}\ \emph {et~al.}(2024)\citenamefont {Delto}, \citenamefont {Duhr}, \citenamefont {Tancredi},\ and\ \citenamefont {Zhu}}]{Delto:2023kqv}%
  \BibitemOpen
  \bibfield  {author} {\bibinfo {author} {\bibfnamefont {M.}~\bibnamefont {Delto}}, \bibinfo {author} {\bibfnamefont {C.}~\bibnamefont {Duhr}}, \bibinfo {author} {\bibfnamefont {L.}~\bibnamefont {Tancredi}},\ and\ \bibinfo {author} {\bibfnamefont {Y.~J.}\ \bibnamefont {Zhu}},\ }\bibfield  {title} {\bibinfo {title} {{Two-Loop QED Corrections to the Scattering of Four Massive Leptons}},\ }\href {https://doi.org/10.1103/PhysRevLett.132.231904} {\bibfield  {journal} {\bibinfo  {journal} {Phys. Rev. Lett.}\ }\textbf {\bibinfo {volume} {132}},\ \bibinfo {pages} {231904} (\bibinfo {year} {2024})},\ \Eprint {https://arxiv.org/abs/2311.06385} {arXiv:2311.06385 [hep-ph]} \BibitemShut {NoStop}%
\bibitem [{\citenamefont {Giroux}\ \emph {et~al.}(2024)\citenamefont {Giroux}, \citenamefont {Pokraka}, \citenamefont {Porkert},\ and\ \citenamefont {Sohnle}}]{Giroux:2024yxu}%
  \BibitemOpen
  \bibfield  {author} {\bibinfo {author} {\bibfnamefont {M.}~\bibnamefont {Giroux}}, \bibinfo {author} {\bibfnamefont {A.}~\bibnamefont {Pokraka}}, \bibinfo {author} {\bibfnamefont {F.}~\bibnamefont {Porkert}},\ and\ \bibinfo {author} {\bibfnamefont {Y.}~\bibnamefont {Sohnle}},\ }\bibfield  {title} {\bibinfo {title} {{The soaring kite: a tale of two punctured tori}},\ }\href {https://doi.org/10.1007/JHEP05(2024)239} {\bibfield  {journal} {\bibinfo  {journal} {JHEP}\ }\textbf {\bibinfo {volume} {05}},\ \bibinfo {pages} {239}},\ \Eprint {https://arxiv.org/abs/2401.14307} {arXiv:2401.14307 [hep-th]} \BibitemShut {NoStop}%
\bibitem [{\citenamefont {Duhr}\ \emph {et~al.}(2024)\citenamefont {Duhr}, \citenamefont {Gasparotto}, \citenamefont {Nega}, \citenamefont {Tancredi},\ and\ \citenamefont {Weinzierl}}]{Duhr:2024bzt}%
  \BibitemOpen
  \bibfield  {author} {\bibinfo {author} {\bibfnamefont {C.}~\bibnamefont {Duhr}}, \bibinfo {author} {\bibfnamefont {F.}~\bibnamefont {Gasparotto}}, \bibinfo {author} {\bibfnamefont {C.}~\bibnamefont {Nega}}, \bibinfo {author} {\bibfnamefont {L.}~\bibnamefont {Tancredi}},\ and\ \bibinfo {author} {\bibfnamefont {S.}~\bibnamefont {Weinzierl}},\ }\bibfield  {title} {\bibinfo {title} {{On the electron self-energy to three loops in QED}},\ }\href {https://doi.org/10.1007/JHEP11(2024)020} {\bibfield  {journal} {\bibinfo  {journal} {JHEP}\ }\textbf {\bibinfo {volume} {11}},\ \bibinfo {pages} {020}},\ \Eprint {https://arxiv.org/abs/2408.05154} {arXiv:2408.05154 [hep-th]} \BibitemShut {NoStop}%
\bibitem [{\citenamefont {Wang}\ \emph {et~al.}(2025)\citenamefont {Wang}, \citenamefont {Wang},\ and\ \citenamefont {Wang}}]{Wang:2024ilc}%
  \BibitemOpen
  \bibfield  {author} {\bibinfo {author} {\bibfnamefont {J.}~\bibnamefont {Wang}}, \bibinfo {author} {\bibfnamefont {X.}~\bibnamefont {Wang}},\ and\ \bibinfo {author} {\bibfnamefont {Y.}~\bibnamefont {Wang}},\ }\bibfield  {title} {\bibinfo {title} {{Analytic decay width of the Higgs boson to massive bottom quarks at order $ {\alpha}_s^3 $}},\ }\href {https://doi.org/10.1007/JHEP03(2025)163} {\bibfield  {journal} {\bibinfo  {journal} {JHEP}\ }\textbf {\bibinfo {volume} {03}},\ \bibinfo {pages} {163}},\ \Eprint {https://arxiv.org/abs/2411.07493} {arXiv:2411.07493 [hep-ph]} \BibitemShut {NoStop}%
\bibitem [{\citenamefont {Forner}\ \emph {et~al.}(2025)\citenamefont {Forner}, \citenamefont {Nega},\ and\ \citenamefont {Tancredi}}]{Forner:2024ojj}%
  \BibitemOpen
  \bibfield  {author} {\bibinfo {author} {\bibfnamefont {F.}~\bibnamefont {Forner}}, \bibinfo {author} {\bibfnamefont {C.}~\bibnamefont {Nega}},\ and\ \bibinfo {author} {\bibfnamefont {L.}~\bibnamefont {Tancredi}},\ }\bibfield  {title} {\bibinfo {title} {{On the photon self-energy to three loops in QED}},\ }\href {https://doi.org/10.1007/JHEP03(2025)148} {\bibfield  {journal} {\bibinfo  {journal} {JHEP}\ }\textbf {\bibinfo {volume} {03}},\ \bibinfo {pages} {148}},\ \Eprint {https://arxiv.org/abs/2411.19042} {arXiv:2411.19042 [hep-th]} \BibitemShut {NoStop}%
\bibitem [{\citenamefont {Schwanemann}\ and\ \citenamefont {Weinzierl}(2025)}]{Schwanemann:2024kbg}%
  \BibitemOpen
  \bibfield  {author} {\bibinfo {author} {\bibfnamefont {N.}~\bibnamefont {Schwanemann}}\ and\ \bibinfo {author} {\bibfnamefont {S.}~\bibnamefont {Weinzierl}},\ }\bibfield  {title} {\bibinfo {title} {{Electroweak double-box integrals for M{\o}ller scattering}},\ }\href {https://doi.org/10.21468/SciPostPhys.18.6.172} {\bibfield  {journal} {\bibinfo  {journal} {SciPost Phys.}\ }\textbf {\bibinfo {volume} {18}},\ \bibinfo {pages} {172} (\bibinfo {year} {2025})},\ \Eprint {https://arxiv.org/abs/2412.07522} {arXiv:2412.07522 [hep-ph]} \BibitemShut {NoStop}%
\bibitem [{\citenamefont {Marzucca}\ \emph {et~al.}(2025)\citenamefont {Marzucca}, \citenamefont {McLeod},\ and\ \citenamefont {Nega}}]{Marzucca:2025eak}%
  \BibitemOpen
  \bibfield  {author} {\bibinfo {author} {\bibfnamefont {R.}~\bibnamefont {Marzucca}}, \bibinfo {author} {\bibfnamefont {A.~J.}\ \bibnamefont {McLeod}},\ and\ \bibinfo {author} {\bibfnamefont {C.}~\bibnamefont {Nega}},\ }\bibfield  {title} {\bibinfo {title} {{Two-Loop Master Integrals for Mixed QCD-EW Corrections to $gg \to H$ Through $\mathcal{O}(\epsilon^2)$}},\ }\href@noop {} {\  (\bibinfo {year} {2025})},\ \Eprint {https://arxiv.org/abs/2501.14435} {arXiv:2501.14435 [hep-th]} \BibitemShut {NoStop}%
\bibitem [{\citenamefont {Becchetti}\ \emph {et~al.}(2025{\natexlab{a}})\citenamefont {Becchetti}, \citenamefont {Coro}, \citenamefont {Nega}, \citenamefont {Tancredi},\ and\ \citenamefont {Wagner}}]{Becchetti:2025rrz}%
  \BibitemOpen
  \bibfield  {author} {\bibinfo {author} {\bibfnamefont {M.}~\bibnamefont {Becchetti}}, \bibinfo {author} {\bibfnamefont {F.}~\bibnamefont {Coro}}, \bibinfo {author} {\bibfnamefont {C.}~\bibnamefont {Nega}}, \bibinfo {author} {\bibfnamefont {L.}~\bibnamefont {Tancredi}},\ and\ \bibinfo {author} {\bibfnamefont {F.~J.}\ \bibnamefont {Wagner}},\ }\bibfield  {title} {\bibinfo {title} {{Analytic two-loop amplitudes for $ q\overline{q}\to \gamma \gamma $ and gg {\textrightarrow} {\ensuremath{\gamma}}{\ensuremath{\gamma}} mediated by a heavy-quark loop}},\ }\href {https://doi.org/10.1007/JHEP06(2025)033} {\bibfield  {journal} {\bibinfo  {journal} {JHEP}\ }\textbf {\bibinfo {volume} {06}},\ \bibinfo {pages} {033}},\ \Eprint {https://arxiv.org/abs/2502.00118} {arXiv:2502.00118 [hep-ph]} \BibitemShut {NoStop}%
\bibitem [{\citenamefont {Becchetti}\ \emph {et~al.}(2025{\natexlab{b}})\citenamefont {Becchetti}, \citenamefont {Dlapa},\ and\ \citenamefont {Zoia}}]{Becchetti:2025oyb}%
  \BibitemOpen
  \bibfield  {author} {\bibinfo {author} {\bibfnamefont {M.}~\bibnamefont {Becchetti}}, \bibinfo {author} {\bibfnamefont {C.}~\bibnamefont {Dlapa}},\ and\ \bibinfo {author} {\bibfnamefont {S.}~\bibnamefont {Zoia}},\ }\bibfield  {title} {\bibinfo {title} {{Canonical differential equations for the elliptic two-loop five-point integral family relevant to tt{\textasciimacron}+jet production at leading color}},\ }\href {https://doi.org/10.1103/zt4w-c1jk} {\bibfield  {journal} {\bibinfo  {journal} {Phys. Rev. D}\ }\textbf {\bibinfo {volume} {112}},\ \bibinfo {pages} {L031501} (\bibinfo {year} {2025}{\natexlab{b}})},\ \Eprint {https://arxiv.org/abs/2503.03603} {arXiv:2503.03603 [hep-th]} \BibitemShut {NoStop}%
\bibitem [{\citenamefont {P{\"o}gel}\ \emph {et~al.}(2025)\citenamefont {P{\"o}gel}, \citenamefont {Teschke}, \citenamefont {Wang},\ and\ \citenamefont {Weinzierl}}]{Pogel:2025bca}%
  \BibitemOpen
  \bibfield  {author} {\bibinfo {author} {\bibfnamefont {S.}~\bibnamefont {P{\"o}gel}}, \bibinfo {author} {\bibfnamefont {T.}~\bibnamefont {Teschke}}, \bibinfo {author} {\bibfnamefont {X.}~\bibnamefont {Wang}},\ and\ \bibinfo {author} {\bibfnamefont {S.}~\bibnamefont {Weinzierl}},\ }\bibfield  {title} {\bibinfo {title} {{The unequal-mass three-loop banana integral}},\ }\href@noop {} {\  (\bibinfo {year} {2025})},\ \Eprint {https://arxiv.org/abs/2507.23594} {arXiv:2507.23594 [hep-th]} \BibitemShut {NoStop}%
\bibitem [{\citenamefont {Duhr}\ \emph {et~al.}(2025{\natexlab{c}})\citenamefont {Duhr}, \citenamefont {Maggio}, \citenamefont {Porkert}, \citenamefont {Semper},\ and\ \citenamefont {Stawinski}}]{Duhr:2025kkq}%
  \BibitemOpen
  \bibfield  {author} {\bibinfo {author} {\bibfnamefont {C.}~\bibnamefont {Duhr}}, \bibinfo {author} {\bibfnamefont {S.}~\bibnamefont {Maggio}}, \bibinfo {author} {\bibfnamefont {F.}~\bibnamefont {Porkert}}, \bibinfo {author} {\bibfnamefont {C.}~\bibnamefont {Semper}},\ and\ \bibinfo {author} {\bibfnamefont {S.~F.}\ \bibnamefont {Stawinski}},\ }\bibfield  {title} {\bibinfo {title} {{Three-loop banana integrals with four unequal masses}},\ }\href@noop {} {\  (\bibinfo {year} {2025}{\natexlab{c}})},\ \Eprint {https://arxiv.org/abs/2507.23061} {arXiv:2507.23061 [hep-th]} \BibitemShut {NoStop}%
\bibitem [{\citenamefont {Frellesvig}(2022)}]{Frellesvig:2021hkr}%
  \BibitemOpen
  \bibfield  {author} {\bibinfo {author} {\bibfnamefont {H.}~\bibnamefont {Frellesvig}},\ }\bibfield  {title} {\bibinfo {title} {{On epsilon factorized differential equations for elliptic Feynman integrals}},\ }\href {https://doi.org/10.1007/JHEP03(2022)079} {\bibfield  {journal} {\bibinfo  {journal} {JHEP}\ }\textbf {\bibinfo {volume} {03}},\ \bibinfo {pages} {079}},\ \Eprint {https://arxiv.org/abs/2110.07968} {arXiv:2110.07968 [hep-th]} \BibitemShut {NoStop}%
\bibitem [{\citenamefont {Chaubey}\ and\ \citenamefont {Sotnikov}(2025)}]{Chaubey:2025adn}%
  \BibitemOpen
  \bibfield  {author} {\bibinfo {author} {\bibfnamefont {E.}~\bibnamefont {Chaubey}}\ and\ \bibinfo {author} {\bibfnamefont {V.}~\bibnamefont {Sotnikov}},\ }\bibfield  {title} {\bibinfo {title} {{Elliptic Leading Singularities and Canonical Integrands}},\ }\href {https://doi.org/10.1103/4fjc-lfnx} {\bibfield  {journal} {\bibinfo  {journal} {Phys. Rev. Lett.}\ }\textbf {\bibinfo {volume} {135}},\ \bibinfo {pages} {101903} (\bibinfo {year} {2025})},\ \Eprint {https://arxiv.org/abs/2504.20897} {arXiv:2504.20897 [hep-th]} \BibitemShut {NoStop}%
\bibitem [{\citenamefont {Chen}\ \emph {et~al.}(2025)\citenamefont {Chen}, \citenamefont {Yang},\ and\ \citenamefont {Zhang}}]{Chen:2025hzq}%
  \BibitemOpen
  \bibfield  {author} {\bibinfo {author} {\bibfnamefont {J.}~\bibnamefont {Chen}}, \bibinfo {author} {\bibfnamefont {L.~L.}\ \bibnamefont {Yang}},\ and\ \bibinfo {author} {\bibfnamefont {Y.}~\bibnamefont {Zhang}},\ }\bibfield  {title} {\bibinfo {title} {{On an approach to canonicalizing elliptic Feynman integrals}},\ }\href@noop {} {\  (\bibinfo {year} {2025})},\ \Eprint {https://arxiv.org/abs/2503.23720} {arXiv:2503.23720 [hep-th]} \BibitemShut {NoStop}%
\bibitem [{\citenamefont {Broedel}\ \emph {et~al.}(2018{\natexlab{a}})\citenamefont {Broedel}, \citenamefont {Duhr}, \citenamefont {Dulat}, \citenamefont {Penante},\ and\ \citenamefont {Tancredi}}]{Broedel:2018iwv}%
  \BibitemOpen
  \bibfield  {author} {\bibinfo {author} {\bibfnamefont {J.}~\bibnamefont {Broedel}}, \bibinfo {author} {\bibfnamefont {C.}~\bibnamefont {Duhr}}, \bibinfo {author} {\bibfnamefont {F.}~\bibnamefont {Dulat}}, \bibinfo {author} {\bibfnamefont {B.}~\bibnamefont {Penante}},\ and\ \bibinfo {author} {\bibfnamefont {L.}~\bibnamefont {Tancredi}},\ }\bibfield  {title} {\bibinfo {title} {{Elliptic symbol calculus: from elliptic polylogarithms to iterated integrals of Eisenstein series}},\ }\href {https://doi.org/10.1007/JHEP08(2018)014} {\bibfield  {journal} {\bibinfo  {journal} {JHEP}\ }\textbf {\bibinfo {volume} {08}},\ \bibinfo {pages} {014}},\ \Eprint {https://arxiv.org/abs/1803.10256} {arXiv:1803.10256 [hep-th]} \BibitemShut {NoStop}%
\bibitem [{\citenamefont {Broedel}\ \emph {et~al.}(2019{\natexlab{a}})\citenamefont {Broedel}, \citenamefont {Duhr}, \citenamefont {Dulat}, \citenamefont {Penante},\ and\ \citenamefont {Tancredi}}]{Broedel:2018qkq}%
  \BibitemOpen
  \bibfield  {author} {\bibinfo {author} {\bibfnamefont {J.}~\bibnamefont {Broedel}}, \bibinfo {author} {\bibfnamefont {C.}~\bibnamefont {Duhr}}, \bibinfo {author} {\bibfnamefont {F.}~\bibnamefont {Dulat}}, \bibinfo {author} {\bibfnamefont {B.}~\bibnamefont {Penante}},\ and\ \bibinfo {author} {\bibfnamefont {L.}~\bibnamefont {Tancredi}},\ }\bibfield  {title} {\bibinfo {title} {{Elliptic Feynman integrals and pure functions}},\ }\href {https://doi.org/10.1007/JHEP01(2019)023} {\bibfield  {journal} {\bibinfo  {journal} {JHEP}\ }\textbf {\bibinfo {volume} {01}},\ \bibinfo {pages} {023}},\ \Eprint {https://arxiv.org/abs/1809.10698} {arXiv:1809.10698 [hep-th]} \BibitemShut {NoStop}%
\bibitem [{\citenamefont {Broedel}\ \emph {et~al.}(2019{\natexlab{b}})\citenamefont {Broedel}, \citenamefont {Duhr}, \citenamefont {Dulat}, \citenamefont {Penante},\ and\ \citenamefont {Tancredi}}]{Broedel:2018rwm}%
  \BibitemOpen
  \bibfield  {author} {\bibinfo {author} {\bibfnamefont {J.}~\bibnamefont {Broedel}}, \bibinfo {author} {\bibfnamefont {C.}~\bibnamefont {Duhr}}, \bibinfo {author} {\bibfnamefont {F.}~\bibnamefont {Dulat}}, \bibinfo {author} {\bibfnamefont {B.}~\bibnamefont {Penante}},\ and\ \bibinfo {author} {\bibfnamefont {L.}~\bibnamefont {Tancredi}},\ }\bibfield  {title} {\bibinfo {title} {{From modular forms to differential equations for Feynman integrals}},\ }in\ \href {https://doi.org/10.1007/978-3-030-04480-0_6} {\emph {\bibinfo {booktitle} {{KMPB Conference}: {Elliptic Integrals, Elliptic Functions and Modular Forms in Quantum Field Theory}}}}\ (\bibinfo {year} {2019})\ pp.\ \bibinfo {pages} {107--131},\ \Eprint {https://arxiv.org/abs/1807.00842} {arXiv:1807.00842 [hep-th]} \BibitemShut {NoStop}%
\bibitem [{\citenamefont {Broedel}\ \emph {et~al.}(2019{\natexlab{c}})\citenamefont {Broedel}, \citenamefont {Duhr}, \citenamefont {Dulat}, \citenamefont {Penante},\ and\ \citenamefont {Tancredi}}]{Broedel:2019hyg}%
  \BibitemOpen
  \bibfield  {author} {\bibinfo {author} {\bibfnamefont {J.}~\bibnamefont {Broedel}}, \bibinfo {author} {\bibfnamefont {C.}~\bibnamefont {Duhr}}, \bibinfo {author} {\bibfnamefont {F.}~\bibnamefont {Dulat}}, \bibinfo {author} {\bibfnamefont {B.}~\bibnamefont {Penante}},\ and\ \bibinfo {author} {\bibfnamefont {L.}~\bibnamefont {Tancredi}},\ }\bibfield  {title} {\bibinfo {title} {{Elliptic polylogarithms and Feynman parameter integrals}},\ }\href {https://doi.org/10.1007/JHEP05(2019)120} {\bibfield  {journal} {\bibinfo  {journal} {JHEP}\ }\textbf {\bibinfo {volume} {05}},\ \bibinfo {pages} {120}},\ \Eprint {https://arxiv.org/abs/1902.09971} {arXiv:1902.09971 [hep-ph]} \BibitemShut {NoStop}%
\bibitem [{\citenamefont {Duhr}\ and\ \citenamefont {Tancredi}(2020)}]{Duhr:2019rrs}%
  \BibitemOpen
  \bibfield  {author} {\bibinfo {author} {\bibfnamefont {C.}~\bibnamefont {Duhr}}\ and\ \bibinfo {author} {\bibfnamefont {L.}~\bibnamefont {Tancredi}},\ }\bibfield  {title} {\bibinfo {title} {{Algorithms and tools for iterated Eisenstein integrals}},\ }\href {https://doi.org/10.1007/JHEP02(2020)105} {\bibfield  {journal} {\bibinfo  {journal} {JHEP}\ }\textbf {\bibinfo {volume} {02}},\ \bibinfo {pages} {105}},\ \Eprint {https://arxiv.org/abs/1912.00077} {arXiv:1912.00077 [hep-th]} \BibitemShut {NoStop}%
\bibitem [{\citenamefont {Weinzierl}(2021)}]{Weinzierl:2020fyx}%
  \BibitemOpen
  \bibfield  {author} {\bibinfo {author} {\bibfnamefont {S.}~\bibnamefont {Weinzierl}},\ }\bibfield  {title} {\bibinfo {title} {{Modular transformations of elliptic Feynman integrals}},\ }\href {https://doi.org/10.1016/j.nuclphysb.2021.115309} {\bibfield  {journal} {\bibinfo  {journal} {Nucl. Phys. B}\ }\textbf {\bibinfo {volume} {964}},\ \bibinfo {pages} {115309} (\bibinfo {year} {2021})},\ \Eprint {https://arxiv.org/abs/2011.07311} {arXiv:2011.07311 [hep-th]} \BibitemShut {NoStop}%
\bibitem [{\citenamefont {Wilhelm}\ and\ \citenamefont {Zhang}(2023)}]{Wilhelm:2022wow}%
  \BibitemOpen
  \bibfield  {author} {\bibinfo {author} {\bibfnamefont {M.}~\bibnamefont {Wilhelm}}\ and\ \bibinfo {author} {\bibfnamefont {C.}~\bibnamefont {Zhang}},\ }\bibfield  {title} {\bibinfo {title} {{Symbology for elliptic multiple polylogarithms and the symbol prime}},\ }\href {https://doi.org/10.1007/JHEP01(2023)089} {\bibfield  {journal} {\bibinfo  {journal} {JHEP}\ }\textbf {\bibinfo {volume} {01}},\ \bibinfo {pages} {089}},\ \Eprint {https://arxiv.org/abs/2206.08378} {arXiv:2206.08378 [hep-th]} \BibitemShut {NoStop}%
\bibitem [{\citenamefont {Adams}\ and\ \citenamefont {Weinzierl}(2018{\natexlab{b}})}]{Adams:2017ejb}%
  \BibitemOpen
  \bibfield  {author} {\bibinfo {author} {\bibfnamefont {L.}~\bibnamefont {Adams}}\ and\ \bibinfo {author} {\bibfnamefont {S.}~\bibnamefont {Weinzierl}},\ }\bibfield  {title} {\bibinfo {title} {{Feynman integrals and iterated integrals of modular forms}},\ }\href {https://doi.org/10.4310/CNTP.2018.v12.n2.a1} {\bibfield  {journal} {\bibinfo  {journal} {Commun. Num. Theor. Phys.}\ }\textbf {\bibinfo {volume} {12}},\ \bibinfo {pages} {193} (\bibinfo {year} {2018}{\natexlab{b}})},\ \Eprint {https://arxiv.org/abs/1704.08895} {arXiv:1704.08895 [hep-ph]} \BibitemShut {NoStop}%
\bibitem [{\citenamefont {Broedel}\ \emph {et~al.}(2019{\natexlab{d}})\citenamefont {Broedel}, \citenamefont {Duhr}, \citenamefont {Dulat}, \citenamefont {Marzucca}, \citenamefont {Penante},\ and\ \citenamefont {Tancredi}}]{Broedel:2019kmn}%
  \BibitemOpen
  \bibfield  {author} {\bibinfo {author} {\bibfnamefont {J.}~\bibnamefont {Broedel}}, \bibinfo {author} {\bibfnamefont {C.}~\bibnamefont {Duhr}}, \bibinfo {author} {\bibfnamefont {F.}~\bibnamefont {Dulat}}, \bibinfo {author} {\bibfnamefont {R.}~\bibnamefont {Marzucca}}, \bibinfo {author} {\bibfnamefont {B.}~\bibnamefont {Penante}},\ and\ \bibinfo {author} {\bibfnamefont {L.}~\bibnamefont {Tancredi}},\ }\bibfield  {title} {\bibinfo {title} {{An analytic solution for the equal-mass banana graph}},\ }\href {https://doi.org/10.1007/JHEP09(2019)112} {\bibfield  {journal} {\bibinfo  {journal} {JHEP}\ }\textbf {\bibinfo {volume} {09}},\ \bibinfo {pages} {112}},\ \Eprint {https://arxiv.org/abs/1907.03787} {arXiv:1907.03787 [hep-th]} \BibitemShut {NoStop}%
\bibitem [{\citenamefont {Broedel}\ \emph {et~al.}(2022)\citenamefont {Broedel}, \citenamefont {Duhr},\ and\ \citenamefont {Matthes}}]{Broedel:2021zij}%
  \BibitemOpen
  \bibfield  {author} {\bibinfo {author} {\bibfnamefont {J.}~\bibnamefont {Broedel}}, \bibinfo {author} {\bibfnamefont {C.}~\bibnamefont {Duhr}},\ and\ \bibinfo {author} {\bibfnamefont {N.}~\bibnamefont {Matthes}},\ }\bibfield  {title} {\bibinfo {title} {{Meromorphic modular forms and the three-loop equal-mass banana integral}},\ }\href {https://doi.org/10.1007/JHEP02(2022)184} {\bibfield  {journal} {\bibinfo  {journal} {JHEP}\ }\textbf {\bibinfo {volume} {02}},\ \bibinfo {pages} {184}},\ \Eprint {https://arxiv.org/abs/2109.15251} {arXiv:2109.15251 [hep-th]} \BibitemShut {NoStop}%
\bibitem [{\citenamefont {Duhr}(2025)}]{Duhr:2025tdf}%
  \BibitemOpen
  \bibfield  {author} {\bibinfo {author} {\bibfnamefont {C.}~\bibnamefont {Duhr}},\ }\bibfield  {title} {\bibinfo {title} {{Modular forms for three-loop banana integrals}},\ }\href {https://doi.org/10.1007/JHEP08(2025)218} {\bibfield  {journal} {\bibinfo  {journal} {JHEP}\ }\textbf {\bibinfo {volume} {08}},\ \bibinfo {pages} {218}},\ \Eprint {https://arxiv.org/abs/2502.15325} {arXiv:2502.15325 [hep-th]} \BibitemShut {NoStop}%
\bibitem [{\citenamefont {Duhr}\ and\ \citenamefont {Maggio}(2025{\natexlab{a}})}]{Duhr:2025ppd}%
  \BibitemOpen
  \bibfield  {author} {\bibinfo {author} {\bibfnamefont {C.}~\bibnamefont {Duhr}}\ and\ \bibinfo {author} {\bibfnamefont {S.}~\bibnamefont {Maggio}},\ }\bibfield  {title} {\bibinfo {title} {{Feynman integrals, elliptic integrals and two-parameter K3 surfaces}},\ }\href {https://doi.org/10.1007/JHEP06(2025)250} {\bibfield  {journal} {\bibinfo  {journal} {JHEP}\ }\textbf {\bibinfo {volume} {06}},\ \bibinfo {pages} {250}},\ \Eprint {https://arxiv.org/abs/2502.15326} {arXiv:2502.15326 [hep-th]} \BibitemShut {NoStop}%
\bibitem [{\citenamefont {Duhr}\ and\ \citenamefont {Maggio}(2025{\natexlab{b}})}]{Duhr:2025ouy}%
  \BibitemOpen
  \bibfield  {author} {\bibinfo {author} {\bibfnamefont {C.}~\bibnamefont {Duhr}}\ and\ \bibinfo {author} {\bibfnamefont {S.}~\bibnamefont {Maggio}},\ }\bibfield  {title} {\bibinfo {title} {{Three-loop banana integrals with three equal masses}},\ }\href@noop {} {\  (\bibinfo {year} {2025}{\natexlab{b}})},\ \Eprint {https://arxiv.org/abs/2511.19245} {arXiv:2511.19245 [hep-th]} \BibitemShut {NoStop}%
\bibitem [{\citenamefont {Frellesvig}\ and\ \citenamefont {Weinzierl}(2024)}]{Frellesvig:2023iwr}%
  \BibitemOpen
  \bibfield  {author} {\bibinfo {author} {\bibfnamefont {H.}~\bibnamefont {Frellesvig}}\ and\ \bibinfo {author} {\bibfnamefont {S.}~\bibnamefont {Weinzierl}},\ }\bibfield  {title} {\bibinfo {title} {{On $\varepsilon$-factorised bases and pure Feynman integrals}},\ }\href {https://doi.org/10.21468/SciPostPhys.16.6.150} {\bibfield  {journal} {\bibinfo  {journal} {SciPost Phys.}\ }\textbf {\bibinfo {volume} {16}},\ \bibinfo {pages} {150} (\bibinfo {year} {2024})},\ \Eprint {https://arxiv.org/abs/2301.02264} {arXiv:2301.02264 [hep-th]} \BibitemShut {NoStop}%
\bibitem [{\citenamefont {Duhr}\ \emph {et~al.}(2025{\natexlab{d}})\citenamefont {Duhr}, \citenamefont {Porkert},\ and\ \citenamefont {Stawinski}}]{Duhr:2024uid}%
  \BibitemOpen
  \bibfield  {author} {\bibinfo {author} {\bibfnamefont {C.}~\bibnamefont {Duhr}}, \bibinfo {author} {\bibfnamefont {F.}~\bibnamefont {Porkert}},\ and\ \bibinfo {author} {\bibfnamefont {S.~F.}\ \bibnamefont {Stawinski}},\ }\bibfield  {title} {\bibinfo {title} {{Canonical differential equations beyond genus one}},\ }\href {https://doi.org/10.1007/JHEP02(2025)014} {\bibfield  {journal} {\bibinfo  {journal} {JHEP}\ }\textbf {\bibinfo {volume} {02}},\ \bibinfo {pages} {014}},\ \Eprint {https://arxiv.org/abs/2412.02300} {arXiv:2412.02300 [hep-th]} \BibitemShut {NoStop}%
\bibitem [{\citenamefont {Bogner}\ \emph {et~al.}(2017)\citenamefont {Bogner}, \citenamefont {Schweitzer},\ and\ \citenamefont {Weinzierl}}]{Bogner:2017vim}%
  \BibitemOpen
  \bibfield  {author} {\bibinfo {author} {\bibfnamefont {C.}~\bibnamefont {Bogner}}, \bibinfo {author} {\bibfnamefont {A.}~\bibnamefont {Schweitzer}},\ and\ \bibinfo {author} {\bibfnamefont {S.}~\bibnamefont {Weinzierl}},\ }\bibfield  {title} {\bibinfo {title} {{Analytic continuation and numerical evaluation of the kite integral and the equal mass sunrise integral}},\ }\href {https://doi.org/10.1016/j.nuclphysb.2017.07.008} {\bibfield  {journal} {\bibinfo  {journal} {Nucl. Phys. B}\ }\textbf {\bibinfo {volume} {922}},\ \bibinfo {pages} {528} (\bibinfo {year} {2017})},\ \Eprint {https://arxiv.org/abs/1705.08952} {arXiv:1705.08952 [hep-ph]} \BibitemShut {NoStop}%
\bibitem [{\citenamefont {Broedel}\ \emph {et~al.}(2018{\natexlab{b}})\citenamefont {Broedel}, \citenamefont {Duhr}, \citenamefont {Dulat},\ and\ \citenamefont {Tancredi}}]{Broedel:2017kkb}%
  \BibitemOpen
  \bibfield  {author} {\bibinfo {author} {\bibfnamefont {J.}~\bibnamefont {Broedel}}, \bibinfo {author} {\bibfnamefont {C.}~\bibnamefont {Duhr}}, \bibinfo {author} {\bibfnamefont {F.}~\bibnamefont {Dulat}},\ and\ \bibinfo {author} {\bibfnamefont {L.}~\bibnamefont {Tancredi}},\ }\bibfield  {title} {\bibinfo {title} {{Elliptic polylogarithms and iterated integrals on elliptic curves. Part I: general formalism}},\ }\href {https://doi.org/10.1007/JHEP05(2018)093} {\bibfield  {journal} {\bibinfo  {journal} {JHEP}\ }\textbf {\bibinfo {volume} {05}},\ \bibinfo {pages} {093}},\ \Eprint {https://arxiv.org/abs/1712.07089} {arXiv:1712.07089 [hep-th]} \BibitemShut {NoStop}%
\bibitem [{Note1()}]{Note1}%
  \BibitemOpen
  \bibinfo {note} {It was suggested in Ref.~\cite {Frellesvig:2023iwr} that they only locally (for $\tau $ in a patch containing the cusp $\protect \mathrm {i}\infty $) satisfy the definition of pure functions in Ref.~\cite {Broedel:2018qkq}, since an $\omega $-form may have an order-$k$ pole in $\tau $ at a finite cusp. This is not a problem in practice, since we can cover the full kinematic space with such local patches with the help of modular transformations.}\BibitemShut {Stop}%
\bibitem [{\citenamefont {P\"ogel}\ \emph {et~al.}(2024)\citenamefont {P\"ogel}, \citenamefont {Wang}, \citenamefont {Weinzierl}, \citenamefont {Wu},\ and\ \citenamefont {Xu}}]{Pogel:2024sdi}%
  \BibitemOpen
  \bibfield  {author} {\bibinfo {author} {\bibfnamefont {S.}~\bibnamefont {P\"ogel}}, \bibinfo {author} {\bibfnamefont {X.}~\bibnamefont {Wang}}, \bibinfo {author} {\bibfnamefont {S.}~\bibnamefont {Weinzierl}}, \bibinfo {author} {\bibfnamefont {K.}~\bibnamefont {Wu}},\ and\ \bibinfo {author} {\bibfnamefont {X.}~\bibnamefont {Xu}},\ }\bibfield  {title} {\bibinfo {title} {{Self-dualities and Galois symmetries in Feynman integrals}},\ }\href {https://doi.org/10.1007/JHEP09(2024)084} {\bibfield  {journal} {\bibinfo  {journal} {JHEP}\ }\textbf {\bibinfo {volume} {09}},\ \bibinfo {pages} {084}},\ \Eprint {https://arxiv.org/abs/2407.08799} {arXiv:2407.08799 [hep-th]} \BibitemShut {NoStop}%
\bibitem [{\citenamefont {Duhr}\ \emph {et~al.}(2025{\natexlab{e}})\citenamefont {Duhr}, \citenamefont {Porkert}, \citenamefont {Semper},\ and\ \citenamefont {Stawinski}}]{Duhr:2024xsy}%
  \BibitemOpen
  \bibfield  {author} {\bibinfo {author} {\bibfnamefont {C.}~\bibnamefont {Duhr}}, \bibinfo {author} {\bibfnamefont {F.}~\bibnamefont {Porkert}}, \bibinfo {author} {\bibfnamefont {C.}~\bibnamefont {Semper}},\ and\ \bibinfo {author} {\bibfnamefont {S.~F.}\ \bibnamefont {Stawinski}},\ }\bibfield  {title} {\bibinfo {title} {{Self-duality from twisted cohomology}},\ }\href {https://doi.org/10.1007/JHEP03(2025)053} {\bibfield  {journal} {\bibinfo  {journal} {JHEP}\ }\textbf {\bibinfo {volume} {03}},\ \bibinfo {pages} {053}},\ \Eprint {https://arxiv.org/abs/2408.04904} {arXiv:2408.04904 [hep-th]} \BibitemShut {NoStop}%
\bibitem [{\citenamefont {Walden}\ and\ \citenamefont {Weinzierl}(2021)}]{Walden:2020odh}%
  \BibitemOpen
  \bibfield  {author} {\bibinfo {author} {\bibfnamefont {M.}~\bibnamefont {Walden}}\ and\ \bibinfo {author} {\bibfnamefont {S.}~\bibnamefont {Weinzierl}},\ }\bibfield  {title} {\bibinfo {title} {{Numerical evaluation of iterated integrals related to elliptic Feynman integrals}},\ }\href {https://doi.org/10.1016/j.cpc.2021.108020} {\bibfield  {journal} {\bibinfo  {journal} {Comput. Phys. Commun.}\ }\textbf {\bibinfo {volume} {265}},\ \bibinfo {pages} {108020} (\bibinfo {year} {2021})},\ \Eprint {https://arxiv.org/abs/2010.05271} {arXiv:2010.05271 [hep-ph]} \BibitemShut {NoStop}%
\bibitem [{\citenamefont {Bezuglov}\ and\ \citenamefont {Onishchenko}(2024)}]{Bezuglov:2023owj}%
  \BibitemOpen
  \bibfield  {author} {\bibinfo {author} {\bibfnamefont {M.~A.}\ \bibnamefont {Bezuglov}}\ and\ \bibinfo {author} {\bibfnamefont {A.~I.}\ \bibnamefont {Onishchenko}},\ }\bibfield  {title} {\bibinfo {title} {{Expansion of hypergeometric functions in terms of polylogarithms with a nontrivial change of variables}},\ }\href {https://doi.org/10.1134/S0040577924060011} {\bibfield  {journal} {\bibinfo  {journal} {Theor. Math. Phys.}\ }\textbf {\bibinfo {volume} {219}},\ \bibinfo {pages} {871} (\bibinfo {year} {2024})},\ \Eprint {https://arxiv.org/abs/2312.06242} {arXiv:2312.06242 [hep-th]} \BibitemShut {NoStop}%
\bibitem [{\citenamefont {Bera}\ and\ \citenamefont {Pathak}(2025)}]{Bera:2024hlq}%
  \BibitemOpen
  \bibfield  {author} {\bibinfo {author} {\bibfnamefont {S.}~\bibnamefont {Bera}}\ and\ \bibinfo {author} {\bibfnamefont {T.}~\bibnamefont {Pathak}},\ }\bibfield  {title} {\bibinfo {title} {{Analytic continuations and numerical evaluation of the Appell F1, F3, Lauricella FD(3) and Lauricella-Saran FS(3) and their application to Feynman integrals}},\ }\href {https://doi.org/10.1016/j.cpc.2024.109386} {\bibfield  {journal} {\bibinfo  {journal} {Comput. Phys. Commun.}\ }\textbf {\bibinfo {volume} {306}},\ \bibinfo {pages} {109386} (\bibinfo {year} {2025})},\ \Eprint {https://arxiv.org/abs/2403.02237} {arXiv:2403.02237 [hep-ph]} \BibitemShut {NoStop}%
\bibitem [{\citenamefont {Bezuglov}\ \emph {et~al.}(2025{\natexlab{a}})\citenamefont {Bezuglov}, \citenamefont {Kniehl}, \citenamefont {Onishchenko},\ and\ \citenamefont {Veretin}}]{Bezuglov:2025xol}%
  \BibitemOpen
  \bibfield  {author} {\bibinfo {author} {\bibfnamefont {M.~A.}\ \bibnamefont {Bezuglov}}, \bibinfo {author} {\bibfnamefont {B.~A.}\ \bibnamefont {Kniehl}}, \bibinfo {author} {\bibfnamefont {A.~I.}\ \bibnamefont {Onishchenko}},\ and\ \bibinfo {author} {\bibfnamefont {O.~L.}\ \bibnamefont {Veretin}},\ }\bibfield  {title} {\bibinfo {title} {{High-precision numerical evaluation of Lauricella functions}},\ }\href {https://doi.org/10.1016/j.nuclphysb.2025.116994} {\bibfield  {journal} {\bibinfo  {journal} {Nucl. Phys. B}\ }\textbf {\bibinfo {volume} {1018}},\ \bibinfo {pages} {116994} (\bibinfo {year} {2025}{\natexlab{a}})},\ \Eprint {https://arxiv.org/abs/2502.03276} {arXiv:2502.03276 [hep-th]} \BibitemShut {NoStop}%
\bibitem [{\citenamefont {Bezuglov}\ \emph {et~al.}(2025{\natexlab{b}})\citenamefont {Bezuglov}, \citenamefont {Kniehl}, \citenamefont {Onishchenko},\ and\ \citenamefont {Veretin}}]{Bezuglov:2025msm}%
  \BibitemOpen
  \bibfield  {author} {\bibinfo {author} {\bibfnamefont {M.~A.}\ \bibnamefont {Bezuglov}}, \bibinfo {author} {\bibfnamefont {B.~A.}\ \bibnamefont {Kniehl}}, \bibinfo {author} {\bibfnamefont {A.~I.}\ \bibnamefont {Onishchenko}},\ and\ \bibinfo {author} {\bibfnamefont {O.~L.}\ \bibnamefont {Veretin}},\ }\bibfield  {title} {\bibinfo {title} {{PrecisionLauricella: Package for numerical computation of Lauricella functions depending on a parameter}},\ }\href {https://doi.org/10.1016/j.cpc.2025.109812} {\bibfield  {journal} {\bibinfo  {journal} {Comput. Phys. Commun.}\ }\textbf {\bibinfo {volume} {316}},\ \bibinfo {pages} {109812} (\bibinfo {year} {2025}{\natexlab{b}})},\ \Eprint {https://arxiv.org/abs/2502.07935} {arXiv:2502.07935 [cs.MS]} \BibitemShut {NoStop}%
\bibitem [{\citenamefont {Morales}\ \emph {et~al.}(2023)\citenamefont {Morales}, \citenamefont {Spiering}, \citenamefont {Wilhelm}, \citenamefont {Yang},\ and\ \citenamefont {Zhang}}]{Morales:2022csr}%
  \BibitemOpen
  \bibfield  {author} {\bibinfo {author} {\bibfnamefont {R.}~\bibnamefont {Morales}}, \bibinfo {author} {\bibfnamefont {A.}~\bibnamefont {Spiering}}, \bibinfo {author} {\bibfnamefont {M.}~\bibnamefont {Wilhelm}}, \bibinfo {author} {\bibfnamefont {Q.}~\bibnamefont {Yang}},\ and\ \bibinfo {author} {\bibfnamefont {C.}~\bibnamefont {Zhang}},\ }\bibfield  {title} {\bibinfo {title} {{Bootstrapping Elliptic Feynman Integrals Using Schubert Analysis}},\ }\href {https://doi.org/10.1103/PhysRevLett.131.041601} {\bibfield  {journal} {\bibinfo  {journal} {Phys. Rev. Lett.}\ }\textbf {\bibinfo {volume} {131}},\ \bibinfo {pages} {041601} (\bibinfo {year} {2023})},\ \Eprint {https://arxiv.org/abs/2212.09762} {arXiv:2212.09762 [hep-th]} \BibitemShut {NoStop}%
\bibitem [{\citenamefont {D'Hoker}\ \emph {et~al.}(2025{\natexlab{a}})\citenamefont {D'Hoker}, \citenamefont {Hidding},\ and\ \citenamefont {Schlotterer}}]{DHoker:2023vax}%
  \BibitemOpen
  \bibfield  {author} {\bibinfo {author} {\bibfnamefont {E.}~\bibnamefont {D'Hoker}}, \bibinfo {author} {\bibfnamefont {M.}~\bibnamefont {Hidding}},\ and\ \bibinfo {author} {\bibfnamefont {O.}~\bibnamefont {Schlotterer}},\ }\bibfield  {title} {\bibinfo {title} {{Constructing polylogarithms on higher-genus Riemann surfaces}},\ }\href {https://doi.org/10.4310/cntp.250531031558} {\bibfield  {journal} {\bibinfo  {journal} {Commun. Num. Theor. Phys.}\ }\textbf {\bibinfo {volume} {19}},\ \bibinfo {pages} {355} (\bibinfo {year} {2025}{\natexlab{a}})},\ \Eprint {https://arxiv.org/abs/2306.08644} {arXiv:2306.08644 [hep-th]} \BibitemShut {NoStop}%
\bibitem [{\citenamefont {Baune}\ \emph {et~al.}(2024)\citenamefont {Baune}, \citenamefont {Broedel}, \citenamefont {Im}, \citenamefont {Lisitsyn},\ and\ \citenamefont {Zerbini}}]{Baune:2024biq}%
  \BibitemOpen
  \bibfield  {author} {\bibinfo {author} {\bibfnamefont {K.}~\bibnamefont {Baune}}, \bibinfo {author} {\bibfnamefont {J.}~\bibnamefont {Broedel}}, \bibinfo {author} {\bibfnamefont {E.}~\bibnamefont {Im}}, \bibinfo {author} {\bibfnamefont {A.}~\bibnamefont {Lisitsyn}},\ and\ \bibinfo {author} {\bibfnamefont {F.}~\bibnamefont {Zerbini}},\ }\bibfield  {title} {\bibinfo {title} {{Schottky{\textendash}Kronecker forms and hyperelliptic polylogarithms}},\ }\href {https://doi.org/10.1088/1751-8121/ad8197} {\bibfield  {journal} {\bibinfo  {journal} {J. Phys. A}\ }\textbf {\bibinfo {volume} {57}},\ \bibinfo {pages} {445202} (\bibinfo {year} {2024})},\ \Eprint {https://arxiv.org/abs/2406.10051} {arXiv:2406.10051 [hep-th]} \BibitemShut {NoStop}%
\bibitem [{\citenamefont {D'Hoker}\ and\ \citenamefont {Schlotterer}(2024)}]{DHoker:2024ozn}%
  \BibitemOpen
  \bibfield  {author} {\bibinfo {author} {\bibfnamefont {E.}~\bibnamefont {D'Hoker}}\ and\ \bibinfo {author} {\bibfnamefont {O.}~\bibnamefont {Schlotterer}},\ }\bibfield  {title} {\bibinfo {title} {{Fay identities for polylogarithms on higher-genus Riemann surfaces}},\ }\href@noop {} {\  (\bibinfo {year} {2024})},\ \Eprint {https://arxiv.org/abs/2407.11476} {arXiv:2407.11476 [hep-th]} \BibitemShut {NoStop}%
\bibitem [{\citenamefont {Baune}\ \emph {et~al.}(2025)\citenamefont {Baune}, \citenamefont {Broedel}, \citenamefont {Im}, \citenamefont {Lisitsyn},\ and\ \citenamefont {Moeckli}}]{Baune:2024ber}%
  \BibitemOpen
  \bibfield  {author} {\bibinfo {author} {\bibfnamefont {K.}~\bibnamefont {Baune}}, \bibinfo {author} {\bibfnamefont {J.}~\bibnamefont {Broedel}}, \bibinfo {author} {\bibfnamefont {E.}~\bibnamefont {Im}}, \bibinfo {author} {\bibfnamefont {A.}~\bibnamefont {Lisitsyn}},\ and\ \bibinfo {author} {\bibfnamefont {Y.}~\bibnamefont {Moeckli}},\ }\bibfield  {title} {\bibinfo {title} {{Higher-genus Fay-like identities from meromorphic generating functions}},\ }\href {https://doi.org/10.21468/SciPostPhys.18.3.093} {\bibfield  {journal} {\bibinfo  {journal} {SciPost Phys.}\ }\textbf {\bibinfo {volume} {18}},\ \bibinfo {pages} {093} (\bibinfo {year} {2025})},\ \Eprint {https://arxiv.org/abs/2409.08208} {arXiv:2409.08208 [hep-th]} \BibitemShut {NoStop}%
\bibitem [{\citenamefont {D'Hoker}\ \emph {et~al.}(2025{\natexlab{b}})\citenamefont {D'Hoker}, \citenamefont {Enriquez}, \citenamefont {Schlotterer},\ and\ \citenamefont {Zerbini}}]{DHoker:2025szl}%
  \BibitemOpen
  \bibfield  {author} {\bibinfo {author} {\bibfnamefont {E.}~\bibnamefont {D'Hoker}}, \bibinfo {author} {\bibfnamefont {B.}~\bibnamefont {Enriquez}}, \bibinfo {author} {\bibfnamefont {O.}~\bibnamefont {Schlotterer}},\ and\ \bibinfo {author} {\bibfnamefont {F.}~\bibnamefont {Zerbini}},\ }\bibfield  {title} {\bibinfo {title} {{Relating flat connections and polylogarithms on higher genus Riemann surfaces}},\ }\href@noop {} {\  (\bibinfo {year} {2025}{\natexlab{b}})},\ \Eprint {https://arxiv.org/abs/2501.07640} {arXiv:2501.07640 [hep-th]} \BibitemShut {NoStop}%
\end{thebibliography}%

\appendix
\widetext

\section{Details of derivations of integrands and letters}

\label{app:details}

In this appendix, we provide detailed derivations for the integrands and symbol letters presented in Section~\ref{sec:can}.

We first give more details about Kronecker-Eisenstein $g$-functions which are essential to the discussions. These functions serve as the generating series for the Kronecker-Eisenstein series $F (z, \alpha, \tau)$,
\begin{equation}
    \label{eq:gdef}
    F (z, \alpha, \tau)
    = \pi \frac{\theta_1^\prime (q) \theta_1 (\pi (z + \alpha), q)}{\theta_1 (\pi z, q) \theta_1 (\pi \alpha, q)}
    = \frac{1}{\alpha} \sum_{k = 0}^{\infty} g^{(k)} (z, \tau) \, {\alpha}^k
    \, ,
\end{equation}
where $q = {\rme}^{\rmi \pi \tau}$ is the \emph{nome}. The functions $g^{(k)}$ are \emph{quasi-periodic} with definite parity:
\begin{equation}
    \label{eq:quasiperiodicparity}
    g^{(k)} (z + 1, \tau)
    = g^{(k)} (z, \tau)
    \, ,
    \quad
    g^{(k)} (z + \tau, \tau)
    = \sum_{i = 0}^{k} \frac{{(- 2 \pi \rmi)}^j}{j !} g^{(k - j)} (z, \tau)
    \, ,
    \quad 
    g^{(k)} (- z, \tau)
    = {(- 1)}^k g^{(k)} (z, \tau)
    \, . 
\end{equation}
All $g^{(k)} (z, \tau)$ have at most simple poles. Specifically, $g^{(1)} (z, \tau)$ has a simple pole at every lattice point, while for $k > 1$, $g^{(k)} (z, \tau)$ has simple poles only at lattice points that do not lie on the real axis. Under modular transformations, $g^{(k)} (z, \tau)$ transforms as a \emph{quasi-modular form} of modular weight-$k$ and \emph{depth}-$k$,
\begin{equation}
    \label{eq:KEgmodtrans}
    g^{(k)} \mleft(\frac{z}{c \tau + d}, \frac{a \tau + b}{c \tau + d}\mright) 
    = {(c \tau + d)}^k \sum_{j = 0}^{k} \frac{(2 \pi \rmi)^j}{j !} {\mleft(\frac{c z}{c \tau + d}\mright)}^j g^{(k - j)} (z, \tau)
    \, .
\end{equation}

To utilize the construction in Eq.~\eqref{eq:canint}, we must express them with Abelian differentials instead of $\Psi$ in integration variable $x$. In order to do this, we first list several useful relations between Abelian differentials and $\Psi$, which we derive based on the identities from Ref.~\cite{Broedel:2017kkb}:
\begin{subequations}
    \label{eq:1formx2z}
    \begin{align}
        \frac{\rmd x}{\sqrt{P_L (x)}}
        & = \omega_1 \, \Psi_0 (0, x) \, \rmd x 
        \, ,
        \\
        \frac{1}{x - e_i} \frac{\sqrt{P_L (e_i)}}{\sqrt{P_L (x)}} \rmd x 
        & = \mleft[\Psi_{- 1} (e_i, x) + 2 g^{(1)} (z_i) \Psi_0 (0, x)\mright] \, \rmd x 
        \, ,
        \\
        \label{eq:z32phipsi}
        \frac{1}{x - e_i} Z_3 (x) \, \rmd x 
        & = 2 \Phi_3 (x) \, \rmd x + \frac{8}{\omega_1} \mleft[\Psi_2 (e_i, x) - \Psi_2 (e_1, x) + g^{(1)} (z_i) \Psi_{- 1} (e_i, x) + 2 g^{(2)} (z_i) \Psi_0 (0, x) - 2 g^{(2)} (z_1) \Psi_0 (0, x)\mright] \, \rmd x 
        \, ,
    \end{align}
\end{subequations}
where we have used the relation $g^{(2)} (z_1) = - \omega_1 \eta_1 $.
The period $\eta_1$, functions $\Phi_3 (x)$ and $Z_3 (x)$ are defined as 
\begin{equation}
    \label{eq:phi3z3def}
    \widetilde{\Phi}_3 (x) 
    \equiv \frac{2}{\sqrt{P_L (x)}} \mleft(- x + \frac{1 + \lambda}{3}\mright)
    \, ,
    \quad 
    \eta_1 
    \equiv \frac{1}{4} \int \limits_{e_2}^{e_3} \, \widetilde{\Phi}_3 (x) \, \rmd x 
    \, ,
    \quad
    \Phi_3 (x)
    \equiv \widetilde{\Phi}_3 (x) - 8 \frac{\eta_1 }{\omega_1} \frac{1}{\sqrt{P_L (x)}} 
    \, ,
    \quad 
    Z_3 (x)
    \equiv \int \limits_{e_4}^{x} \, \Phi_3 (x^\prime) \, \rmd x^\prime
    \, .
\end{equation}

With Eq.~\eqref{eq:1formx2z}, we can easily express the canonical integrands of the first and third kinds in Eq.~\eqref{eq:canint} exactly in terms of Abelian differentials 
\begin{subequations}
    \label{eq:canint13}
    \begin{align}
        \phi_1 
        & = \pi \, \de \mleft(
            \begin{smallmatrix}
                0 
                \\
                0
            \end{smallmatrix}; 
            x, \lambda
            \mright)
        = \pi \, \Psi_0 (0, x) \, \rmd x 
        = \frac{\pi}{\omega_1} \frac{\rmd x}{\sqrt{P_L (x)}}
        \, ,
        \\
        \phi_{i - 3}
        & = \beta_i \, \de \mleft(
            \begin{smallmatrix}
                - 1 
                \\
                e_i
            \end{smallmatrix}; 
            x, \lambda
            \mright)
        = \beta_i \, \Psi_{- 1} (e_i, x) \, \rmd x
        = \beta_i \mleft[\frac{1}{x - e_i} \frac{\sqrt{P_L (e_i)}}{\sqrt{P_L (x)}} - \frac{2}{\omega_1} g^{(1)} (z_i, \tau) \frac{1}{\sqrt{P_L (x)}}\mright] \, \rmd x
        \, ,
        \quad 
        (i = 5, \cdots, n + 1)
        \, .
    \end{align}
\end{subequations}
For canonical integrands of the second kind, we first observe in Eq.~\eqref{eq:z32phipsi} that while $\Phi_3 (x) \, \rmd x$ is an Abelian differential, the term involving $Z_3 (x)$ is not. Consequently, we must find a linear combination of $Z_3 (x)$ terms that reduces to an Abelian differential. We first consider the integral
\begin{equation}
    \label{eq:canint2}
    \int \limits_\mathcal{C} \, \bar{u}_L (x) \, \Phi_3 (x) \, \rmd x 
    = - \int \limits_\mathcal{C} \, \bar{u}_L (x) \, Z_3 (x) \, \dlog \mleft[\bar{u}_L (x)\mright]
    = \varepsilon \sum_{i = 2}^{n + 1} \beta_i \int \limits_\mathcal{C} \, \bar{u}_L (x) \frac{1}{x - e_i} Z_3 (x) \, \rmd x 
    \, .
\end{equation}
With the relations in Eq.~\eqref{eq:1formx2z} and $\beta_1 = - \sum_{i = 2}^{n + 1} \beta_i$, we can express the canonical integrands of the second kind in Eq.~\eqref{eq:canint} as
\begin{equation}
    \label{eq:canint2ibp}
    \phi_{n - 2} 
    = \frac{1}{\pi} \sum_{i = 1}^{n + 1} \beta_i \, \de \mleft(
            \begin{smallmatrix}
                2 
                \\
                e_i
            \end{smallmatrix}; 
            x, \lambda
            \mright)
    = \frac{1}{\pi} \sum_{i = 2}^{n + 1} \beta_i \, \Psi_2 (e_i, x) \, \rmd x 
    \sim \frac{1 + 2 \beta_1 \varepsilon}{8 \pi \varepsilon} \omega_1 \, \Phi_3 (x) \, \rmd x - \frac{1}{\pi} \sum_{i = 1}^{n + 1} \beta_i \, \Upsilon (e_i, x) \, \rmd x 
    \, ,
\end{equation}
where $\sim$ means integral-level equality under IBP. The functions $\Upsilon$ are defined as
\begin{equation}
    \label{eq:upsilon}
    \Upsilon (e_i, x)
    \equiv 
    \begin{cases}
        2 g^{(2)} (z_1, \tau) \, \Psi_0 (0, x) \, ,  & i = 1
        \\
        g^{(1)} (z_i, \tau) \, \Psi_{- 1} (e_i, x) + 2 g^{(2)} (z_i, \tau) \, \Psi_0 (0, x) \, , & i = 2, \cdots, n + 1
    \end{cases}
    \, .
\end{equation}
The only subtleties come from $\Upsilon (e_i, x)$-terms for $i = 1, 2, 3, 4$, which are 
\begin{equation}
    \label{eq:canint3sp} 
    \Psi_{- 1} (e_{1, 4}, x) \, \rmd x  
    = 0 
    \, ,
    \quad 
    \Psi_{- 1} (e_{2, 3}, x) \, \rmd x 
    = 2 \rmi \pi \, \Psi_0 (0, x) \, \rmd x 
    \, ,
    \quad 
    g^{(1)} (z_{2, 3}, \tau) 
    = - \rmi \pi 
    \, ,
    \quad 
    g^{(1)} (z_4, \tau)
    = 0 
    \, .
\end{equation}

We next derive the symbol letters. The $\varepsilon^0$-order terms in Eq.~\eqref{eq:epsexp} can be directly expressed in terms of $\widetilde{\Gamma}$ with Eq.~\eqref{eq:purekernel},
\begin{subequations}
    \label{eq:eps0gamma}
    \begin{align}
        \int \limits_\mathcal{C} \, \widetilde{\phi}_0
        & = \int \limits_{0}^{z^\star} \, \pi \, \rmd z
        = \pi \, \widetilde{\Gamma} 
        \mleft(
            \begin{smallmatrix}
                0
                \\
                0
            \end{smallmatrix}; z^\star, \tau
        \mright)
        \, ,
        \\
        \int \limits_\mathcal{C} \, \widetilde{\phi}_i 
        & = \beta_i \int \limits_{0}^{z^\star} \, \mleft[g^{(1)} (z - z_i, \tau) - g^{(1)} (z + z_i, \tau)\mright] \, \rmd z
        = \beta_i 
        \mleft[ 
        \widetilde{\Gamma} 
        \mleft(
            \begin{smallmatrix}
                1
                \\
                z_i^+
            \end{smallmatrix}; z^\star, \tau
        \mright) 
        - \widetilde{\Gamma}
        \mleft(
            \begin{smallmatrix}
                1 
                \\
                z_i^-
            \end{smallmatrix}; z^\star, \tau
        \mright)
        \mright]
        \, ,
        \quad 
        (i = 1, \cdots, n + 1)
        \, ,
        \\
        \int \limits_\mathcal{C} \, \widetilde{\phi}_{n + 2} 
        & = \frac{1}{\pi} \sum_{i = 1}^{n + 1} \beta_i \int \limits_{0}^{z^\star} \, \mleft[g^{(2)} (z - z_i, \tau) + g^{(2)} (z + z_i, \tau)\mright] \, \rmd z
        = \frac{1}{\pi} \sum_{i = 1}^{n + 1} \beta_i 
        \mleft[
            \widetilde{\Gamma} 
            \mleft(
                \begin{smallmatrix}
                    2 
                    \\
                    z_i^+
                \end{smallmatrix}; z^\star, \tau
            \mright)
            + \widetilde{\Gamma}
            \mleft(
                \begin{smallmatrix}
                    2
                    \\
                    z_i^-
                \end{smallmatrix}; z^\star, \tau 
            \mright)
        \mright]
        \, ,
    \end{align}
\end{subequations}
where the contour $\mathcal{C}$ is chosen as $[e_1, e_4]$ which is mapped to $[0, z^\star]$ with $z^\star=1 / 2$. It is easy to verify that the total differentials of the $\varepsilon^0$-order terms vanish as expected. For the $\varepsilon^1$-order terms, we first express $\log (x - e_i)$'s in terms of $\widetilde{\Gamma}$,
\begin{align}
    \label{eq:log2gamma}
    \log (x - e_i) 
    & = \lim_{\delta \to 0} \mleft[\int \limits_{\frac{1}{\delta}}^{x} \,  \frac{1}{x^\prime - e_i} \, \rmd x^\prime - \log \delta\mright]
    \nonumber
    \\
    & = \lim_{\delta \to 0} \mleft\{\int \limits_{\epsilon(\delta)}^{z} \, \mleft[g^{(1)} (z^\prime - z_i, \tau) + g^{(1)} (z^\prime + z_i, \tau) - 2 g^{(1)} (z^\prime, \tau)\mright] \rmd z^\prime - \log \delta\mright\}
    \nonumber
    \\
    & = \lim_{\delta \to 0} 
    \mleft[
        \widetilde{\Gamma}
        \mleft(
            \begin{smallmatrix}
                1
                \\
                z_i^+
            \end{smallmatrix}; z, \tau 
        \mright) 
        + \widetilde{\Gamma}
        \mleft(
            \begin{smallmatrix}
                1
                \\
                z_i^-
            \end{smallmatrix}; z, \tau 
        \mright)
        - 2 \widetilde{\Gamma}
        \mleft(
            \begin{smallmatrix}
                1
                \\
                0
            \end{smallmatrix}; z, \tau 
        \mright) + 2 \log \epsilon (\delta) - \log \delta
    \mright]
    \, .
\end{align}
Note we ignore potential constant terms irrelevant to the total differentials and the corresponding differential equations here and in the following. The relation between $\epsilon (\delta)$ and $\delta$ can be derived from the Abel's map in Eq.~\eqref{eq:abelmap},
\begin{equation}
    \label{eq:epsdelta}
    \epsilon (\delta)
    = \frac{1}{\omega_1} \int \limits_{\infty}^{\frac{1}{\delta}} \, \frac{\rmd x}{\sqrt{P_L (x)}}
    = \frac{1}{\omega_1} \int \limits_{0}^{\delta} \, \frac{\rmd t}{\sqrt{t (1 - \lambda t) (1 - t)}}
    = \frac{1}{2 \omega_1} \sqrt{\delta} + {\cal O} \mleft(\delta^{3 / 2}\mright)
    \, .
\end{equation}
Plugging it into Eq.~\eqref{eq:log2gamma}, we have 
\begin{equation}
    \label{eq:log2gammalog}
    \log (x - e_i) 
    = \widetilde{\Gamma}
        \mleft(
            \begin{smallmatrix}
                1
                \\
                z_i^+
            \end{smallmatrix}; z, \tau 
        \mright) 
        + \widetilde{\Gamma}
        \mleft(
            \begin{smallmatrix}
                1
                \\
                z_i^-
            \end{smallmatrix}; z, \tau 
        \mright)
        - 2 \widetilde{\Gamma}
        \mleft(
            \begin{smallmatrix}
                1
                \\
                0
            \end{smallmatrix}; z, \tau 
        \mright) - 2 \log \omega_1
        \, .
\end{equation}

We can then express the $\varepsilon^1$-order term in terms of $\widetilde{\Gamma}$ by inserting Eq.~\eqref{eq:log2gammalog} into Eq.~\eqref{eq:epsexp}. After some algebra, we have
\begin{subequations}
    \label{eq:eps1gamma}
    \begin{align}
        \sum_{i = 2}^{n + 1} \beta_i \int \limits_\mathcal{C} \, \widetilde{\phi}_0 \log (x - e_i) 
        & = \sum_{i = 1}^{n + 1} \pi \beta_i 
        \mleft[
            \widetilde{\Gamma} 
            \mleft(
                \begin{smallmatrix}
                    0 & 1 
                    \\
                    0 & z_i^+
                \end{smallmatrix}; z^\star, \tau 
            \mright)
            + \widetilde{\Gamma} 
            \mleft(
                \begin{smallmatrix}
                    0 & 1 
                    \\
                    0 & z_i^-
                \end{smallmatrix}; z^\star, \tau 
            \mright)
        \mright]
        + 2 \pi \beta_1 \log \omega_1 \, 
        \widetilde{\Gamma} 
        \mleft(
            \begin{smallmatrix}
                0
                \\
                0
            \end{smallmatrix}; z^\star, \tau
        \mright)
        \, ,
        \\
        \sum_{j = 2}^{n + 1} \beta_j \int \limits_\mathcal{C} \, \widetilde{\phi}_i \log (x - e_j) 
        & = \beta_i \sum_{j = 1}^{n + 1} \beta_j 
        \mleft[
            \widetilde{\Gamma} 
            \mleft(
                \begin{smallmatrix}
                    1 & 1 
                    \\
                    z_i^+ & z_j^+
                \end{smallmatrix}; z^\star, \tau 
            \mright)
            + \widetilde{\Gamma} 
            \mleft(
                \begin{smallmatrix}
                    1 & 1 
                    \\
                    z_i^+ & z_j^-
                \end{smallmatrix}; z^\star, \tau 
            \mright)
            - \widetilde{\Gamma} 
            \mleft(
                \begin{smallmatrix}
                    1 & 1 
                    \\
                    z_i^- & z_j^+
                \end{smallmatrix}; z^\star, \tau 
            \mright)
            - \widetilde{\Gamma} 
            \mleft(
                \begin{smallmatrix}
                    1 & 1 
                    \\
                    z_i^- & z_j^-
                \end{smallmatrix}; z^\star, \tau 
            \mright)
        \mright]
        \nonumber
        \\
        & + 2 \beta_1 \beta_i \log \omega_1 \,
        \mleft[
            \widetilde{\Gamma} 
            \mleft(
                \begin{smallmatrix}
                    1
                    \\
                    z_i^+
                \end{smallmatrix}; z^\star, \tau
            \mright) 
            - \widetilde{\Gamma}
            \mleft(
                \begin{smallmatrix}
                    1 
                    \\
                    z_i^-
                \end{smallmatrix}; z^\star, \tau
            \mright)
        \mright]
        \, ,
        \quad 
        (i = 1, \cdots, n + 1)
        \, ,
        \\
        \sum_{i = 2}^{n + 1} \beta_i \int \limits_\mathcal{C} \, \widetilde{\phi}_{n + 2} \log (x - e_i) 
        & = \sum_{i, j = 1}^{n + 1} \frac{\beta_i \beta_j}{\pi} 
        \mleft[
            \widetilde{\Gamma} 
            \mleft(
                \begin{smallmatrix}
                    2 & 1 
                    \\
                    z_i^+ & z_j^+
                \end{smallmatrix}; z^\star, \tau 
            \mright)
            + \widetilde{\Gamma} 
            \mleft(
                \begin{smallmatrix}
                    2 & 1 
                    \\
                    z_i^+ & z_j^-
                \end{smallmatrix}; z^\star, \tau 
            \mright)
            + \widetilde{\Gamma} 
            \mleft(
                \begin{smallmatrix}
                    2 & 1 
                    \\
                    z_i^- & z_j^+
                \end{smallmatrix}; z^\star, \tau 
            \mright)
            + \widetilde{\Gamma} 
            \mleft(
                \begin{smallmatrix}
                    2 & 1 
                    \\
                    z_i^- & z_j^-
                \end{smallmatrix}; z^\star, \tau 
            \mright)
        \mright]
        \nonumber
        \\
        & + \frac{2}{\pi} \beta_1 \log \omega_1 \sum_{i = 1}^{n + 1} \beta_i \,
        \mleft[
            \widetilde{\Gamma} 
            \mleft(
                \begin{smallmatrix}
                    2
                    \\
                    z_i^+
                \end{smallmatrix}; z^\star, \tau
            \mright) 
            + \widetilde{\Gamma}
            \mleft(
                \begin{smallmatrix}
                    2
                    \\
                    z_i^-
                \end{smallmatrix}; z^\star, \tau
            \mright)
        \mright]
        \, .
    \end{align}
\end{subequations}

In order to obtain the total derivatives of the $\varepsilon^1$-order terms, we need the total differential of $\widetilde{\Gamma}$, which is
\begin{equation}
  \label{eq:empldiff}
  \begin{aligned}
    \rmd \widetilde{\Gamma} \mleft(
      A_1 \cdots A_k; z, \tau
    \mright)
    & = \sum_{p = 1}^{k - 1} {(- 1)}^{n_{p + 1}} {\pi}^{n_p + n_{p + 1} - 1} \widetilde{\Gamma} 
    \mleft(
      A_1 \cdots A_{p - 1} 
      \begin{smallmatrix}
        0 
        \\
        0
      \end{smallmatrix}
      A_{p + 2} \cdots A_k;
      z, \tau
    \mright) \, \omega^{(n_p + n_{p + 1} + 1)}_{p, p + 1}
    \\
    & + \sum_{p = 1}^{k} \sum_{r = 0}^{n_p + 1} {\pi}^{n_p - r - 1} 
    \mleft[\mleft(
      \begin{array}{c}
        n_{p - 1} + r - 1
        \\
        n_{p - 1} - 1
      \end{array}
    \mright) \, \widetilde{\Gamma} 
    \mleft(
      A_1 \cdots A^{[r]}_{p - 1} \hat{A}_p A_{p + 1} \cdots A_k; 
      z, \tau
    \mright) \, \omega^{(n_p - r + 1)}_{p, p - 1} 
    \mright. 
    \\
    & - \mleft.\mleft(
      \begin{array}{c}
        n_{p + 1} + r - 1
        \\
        n_{p + 1} - 1
      \end{array}
    \mright) \, \widetilde{\Gamma} 
    \mleft(
      A_1 \cdots A_{p - 1} \hat{A}_p A^{[r]}_{p + 1} \cdots A_k; 
      z, \tau
    \mright) \, \omega^{(n_p - r + 1)}_{p, p + 1}
    \mright]  
    \, ,
  \end{aligned}
\end{equation}
where we introduce the shorthand notations
\begin{equation}
  \label{eq:shorthand}
  \omega^{(n)}_{i, j}
  \equiv \omega^{(n)} (z_j - z_i, \tau)
  \, ,
  \quad
  A^{[r]}_i 
  \equiv \mleft(
    \begin{array}{c}
      n_i + r 
      \\
      z_i
    \end{array}
  \mright)
  \, ,
  \quad 
  (z_0, z_{k + 1})
  = (z, 0)
  \, ,
  \quad 
  (n_0, n_{k + 1})
  = (0, 0)
  \, .
\end{equation}
Thus, we can obtain the total differential of the $\varepsilon^1$-order terms in Eq.~\eqref{eq:eps1gamma} by setting $z^\star = 1 / 2$ and applying Eq.~\eqref{eq:empldiff} straightforwardly. After some algebra, we arrive at the final result in Eq.~\eqref{eq:Atilde}.

With the extended canonical integrands in Eq.~\eqref{eq:extcanint} in terms of $\de$-forms~\eqref{eq:purekernel}, and the transformations properties of $g^{(k)} (z, \tau)$ in Eq.~\eqref{eq:KEgmodtrans}, we can derive the additional fiber transformation for the canonical basis under modular transformations,
\begin{subequations}
    \label{eq:puremodtrans}
    \begin{align}
        \widetilde{\phi}_0
        & \mapsto 
        \frac{1}{c \tau + d} \widetilde{\phi}_0
        \, ,
        \\
        \widetilde{\phi}_{i}
        & \mapsto 
        \widetilde{\phi}_{i} - 4 \rmi \frac{c \beta_i z_i}{c \tau + d} \widetilde{\phi}_0 
        \, ,
        \quad 
        (i = 1, \cdots, n + 1)
        \, ,
        \\
        \widetilde{\phi}_{n + 2}
        & \mapsto 
        (c \tau + d) \widetilde{\phi}_{n + 2} - 2 \rmi c \sum_{i = 1}^{n + 1} z_i \, \widetilde{\phi}_{i} - 4 \frac{{c}^2}{c \tau + d} \sum_{i = 1}^{n + 1} \beta_i \, {z_i}^2 \, \widetilde{\phi}_0 + 2 \rmi c \sum_{i = 2}^{n + 1} \beta_i \frac{z}{x - e_i} \rmd x 
        \, .
    \end{align}
\end{subequations}

The integrands of the first and the third kinds exhibit simple behaviors under modular transformations, while the second kind involves a more complicated structure, with $z / (x - e_i) \, \rmd x$-terms. Fortunately, we can remove these terms with integral-level relations. Specifically, we consider the integral
\begin{equation}
    \label{eq:pure2modtransibp}
    \int \limits_\mathcal{C} \, \bar{u}_L (x) \, \widetilde{\phi}_0 
    = - \pi \int \limits_\mathcal{C} \, \bar{u}_L (x) \, z \, \dlog \bar{u}_L (x) 
    = \pi \varepsilon \sum_{i = 2}^{n + 1} \beta_i \int \limits_\mathcal{C} \, \bar{u}_L (x) \frac{z}{x - e_i} \rmd x
    \, .
\end{equation}
Thus, we have the integral-level relation 
\begin{equation}
    \label{eq:zpsi1rel}
    \sum_{i = 2}^{n + 1} \beta_i \frac{z}{x - e_i} \rmd x
    \sim \frac{1}{\pi \varepsilon} \widetilde{\phi}_0 
    \, .
\end{equation}
Plugging Eq.~\eqref{eq:zpsi1rel} into Eq.~\eqref{eq:puremodtrans}, we have the additional fiber transformation under modular transformations in Eq.~\eqref{eq:fibermodtranscanext}.

\end{document}